\documentclass[12pt,letterpaper,magyar,english]{article}
\usepackage[T1]{fontenc}
\usepackage[latin2,latin9]{inputenc}
\usepackage{amsmath}
\usepackage{graphicx}
\usepackage{amssymb}

\makeatletter

\providecommand{\tabularnewline}{\\}

\newcommand{\lyxaddress}[1]{
\par {\raggedright #1
\vspace{1.4em}
\noindent\par}
}

\usepackage{a4wide}

\newcommand\prd[3]   { {{\it Phys.\ Rev.\ }{\bf D #1} (#2) #3}}

\makeatother

\usepackage{babel}

\begin{document}

\title{\textbf{\Large Fermion Condensate as Higgs substitute}}

\author{G. Cynolter and E. Lendvai}

\date{$\phantom{.}$}

\maketitle

\lyxaddress{\textit{Theoretical Physics Research Group of Hungarian Academy of
Sciences, Eötvös University, Budapest, 1117 Pázmány Péter sétány 1/A,
Hungary}\\
\textit{ }}
\begin{abstract}
We propose and analyze an alternative model of dynamical electroweak
symmetry breaking. In the Standard Model of electroweak interactions
the elementary Higgs field and the Higgs sector are replaced by vector-like
fermions and their interactions. The new fermions are a weak doublet
and a singlet. They have kinetic terms with covariant derivatives
and gauge invariant four-fermion interactions. The model is a low
energy effective one with a natural cutoff in the TeV regime. Due
to the quartic fermion couplings the new fermions form condensates.
The new fermions mix in one condensate and the mixing breaks the electroweak
symmetry. The condensates contribute to the masses of the new femions,
which may or may not have mass terms in the original Lagrangian. Gap
equations are derived for the masses of the new fermions and the conditions
are presented for mass generations and electroweak symmetry breaking.
In the spectrum there are two neutral fermions and a charged one with
mass between the neutral ones. The new sector can be described by
three parameters, these are the two neutral masses and the mixing
angle. These parameters are further constrained by the unitarity of
two particle scattering amplitudes, providing an upper bound for the
lighter neutral mass depending on the cutoff of the model. The standard
chiral fermions get there masses via interactions with the condensing
new fermions, but there is no mixing between the standard and the
new fermions. There is an effective composite scalar in the model
at low energies, producing the weak gauge boson masses in effective
interactions. The $\rho$ parameter is one at leading order. The model
can be constrained by one-loop oblique corrections. The Peskin-Takeuchi
S and T parameters are calculated in the model. The parameters of
the model are only slightly constrained, the T parameter requires
the new neutral fermion masses not to be very far from each other,
allowing higher mass difference for higher masses and smaller mixing.
The S parameter gives practically no constraints on the masses. The
new fermions can give positive contributions to T allowing for a heavy
Higgs in the precision electroweak tests. It is shown that the new
fermions will be copiously produced at the next generation of linear
colliders and cross sections are presented for the Large Hadron Collider.
An additional nice feature of the model is that the lightest new neutral
fermion is an ideal and natural dark matter candidate. 
\end{abstract}

\section{Introduction}

The Standard Model of particle physics successfully describes known
collider experiments reaching the permille level in case of some observables.
The only missing particle of the Standard Model is the elementary
Higgs boson. In the minimal Standard Model a weak doublet (hypercharge
Y=1) scalar field is postulated with an \textit{ad hoc} scalar potential
to trigger electroweak symmetry breaking. This provides a very economical
and simple description. Three Goldstone Bosons are eaten up by the
$W^{\pm},\, Z$ gauge bosons providing their correct masses, but the
remaining single CP-even neutral Higgs scalar has evaded the experimental
discovery so far. There exist experimental constraints on the mass
of the Higgs boson. The LEP2 experiment has put a lower bound $M_{H}>114.4$
GeV \cite{pdg} and there is an exclusion window from the combined
D0 and CDF measurements at the Tevatron \cite{tevatron10} between
158 and 175 GeV. The precision data favour a light Higgs with a central
value below the direct LEP2 bound. Including the results of the direct
searches both at LEP2 and the Tevatron the upper limit is driven to
$M_{H}\leq147$ at 95 \% C.L. from electroweak precision tests \cite{pdg}.
The Gfitter group has arrived at similar upper bounds $M_{H}\leq159$
GeV (155 GeV) with or without the information of the direct Higgs
searches \cite{gfitter}.

Beside the missing experimental discovery, theories with elementary
scalars are burdened with theoretical problems, like triviality and
the most severe gauge hierarchy problem. Elementary scalars are unstable
against radiative corrections and without fine tuning the Standard
Model must be cut off at few TeV.

There are mainly two ways to solve these problems in particle physics,
either impose new symmetries to protect the scalars or eliminate elementary
scalars from the theory. 

Supersymmetry is the number one candidate for beyond the Standard
Model physics, it protects the quadratically unstable Higgs mass,
the contribution of the superpartners cancel each other. The Minimal
Supersymmetric Standard Model is very attractive considering that
electroweak symmetry breaking is triggered radiatively, there are
ideal dark matter candidates and gauge couplings unify better in supersymmetric
Grand Unified Theories than in standard GUTs. However supersymmetric
theories involve a huge parameter space, all known particles are doubled
and no satisfactory mechanism has been worked out for supersymmetry
breaking. None of the predicted new superpartners have been found
in any of the experiments and supersymmetry may start to lose it's
appeal. Another shortcoming is that with no discovery the superpartner
masses and the scale of supersymmetry breaking are pushed higher and
higher reformulating the fine tuning problem at a percent level. 

There are strong indications, expectations and a {}``no lose theorem''
that the LHC will reveal the physics of electroweak symmetry breaking.
Either the LHC will find one or more Higgs bosons, it could be the
Standard Model one or a scalar coming from an extended Higgs sector
like the MSSM or the LHC will discover some sign of new, possibly
strong dynamics that unitarizes the scattering of longitudinal gauge
bosons in the TeV regime. These observations motivate to study alternative
models of electroweak symmetry breaking without elementary scalars.

The other main solution to the hierarchy problem employs the mechanism
of dynamical symmetry breaking. The original technicolor idea \cite{Weinberg,Weinberg2,Susskind}
of fermion condensation is already more than thirty years old, it
is based on real phenomena of QCD. Technicolor still gives motivation
for new research, see a recent review \cite{Hillpr}, Chivukula et
al. in \cite{pdg} and references therein. New chiral fermions are
postulated which are charged under the new technicolor gauge group,
the new interaction becomes strong condensing the techni-fermions
charged under the weak $SU_{L}(2)$. To provide fermion masses extended
technicolor gauge interactions (ETC) \cite{etc,etc2} must be included.
The tension between sizeable quark masses and avoiding flavor changing
neutral currents led to introduce walking, near conformal dynamics
\cite{holdom,bando}. These ideas and the phase diagram of strongly
interacting models triggered activity in lattice studies \cite{latticeconf},
and further new technicolor models were constructed based on adjoint
or two index symmetric representations of the new fermions \cite{sannino}.
The heavy top quark is natural in top condensate models \cite{t1,t2},
and there are extra dimensional realizations, too \cite{t3,t4}. 

Inspired by discretized higher dimensional theories \textquotedbl{}little
Higgs\textquotedbl{} \cite{lH} models provide a new class of composite
Higgs models, and they attracted considerable interest solving the
{}``little hierarchy problem'' \cite{kishier} allowing to raise
the cutoff of the theory up to 10 TeV without excessive fine tuning
\cite{lh1,lh}. Little Higgs models realize the old idea that the
Higgs is a pseudo Goldstone boson of some spontaneously broken global
symmetry \cite{pshiggs}. Contrary to supersymmetric models divergent
fermion (boson) loops cancel fermion (boson) loops. Little Higgs models
still require large fine tuning unless they posses custodial symmetry
at the price of highly extended gauge groups. There are various models
where the Higgs is composite \cite{comp}, the idea was recently realized
in extra dimensions \cite{compmin}. Higgsless models \cite{csaki}
do not utilize a scalar Higgs boson, but using the AdS/CFT correspondence
these are extra dimensional \textquotedbl{}duals\textquotedbl{} of
walking technicolor theories.

In this chapter we present a recently proposed alternative symmetry
breaking model of electroweak interactions \cite{fcm}. The complete
symmetry breaking sector is built from a new doublet and a singlet
vector-like fermions, the Higgs is a composite state of the new fermions.
Using vector-like fermions is advantageous compared to chiral ones
as the constraints from precision electroweak measurements are much
weaker. Vector-like fermions appear in several extensions of the Standard
Model. They are present in extra dimensional models with bulk fermions
e.g \cite{UED}, in little Higgs theories \cite{lH,lh1,lh}, in models
of so called improved naturalness consistent with a heavy Higgs scalar
\cite{improved}, in simple fermionic models of dark matter \cite{darkmatter,dmhiggs},
in dynamical models of supersymmetry breaking using gauge medation,
topcolor models \cite{top}. Vector-like fermions were essential ingredients
of our proposal, in which a nontrivial condensate of new vector-like
fermions breaks the electroweak symmetry and provides masses for the
standard particles \cite{fcm}. 

In the Fermion Condensate Model the Higgs sector is replaced by the
interactions of a new doublet $\Psi_{D}=\left(\begin{array}{c}
\Psi_{D}^{+}\\
\Psi_{D}^{0}\end{array}\right)$ and a singlet $\Psi_{S}$ hypercharge 1 vector-like (non-chiral)
fermion field. After electroweak symmetry breaking $\Psi_{D}^{+}$
field corresponds to a positively charged particle and $\Psi_{D}^{0}$
to a neutral one. The new fermions are postulated to have effective
non-renormalizable four-fermion interactions and the model is a low
energy effective one, valid up to some intrinsic, physical cutoff,
that is not be taken to infinity. Therefore we are not forced to add
additional terms to calculate at lowest orders following \cite{njl},
including extra terms will define a different model. The ultraviolet
completion of the model is not yet specified, but as usual the four-fermion
terms are expected to originated from some spontanously broken gauge
interactions. The key point is that the four-fermion interactions
become strong at low energies and generate condensates of the new
fermions including a mixed condensate of $\Psi_{D}$ and $\bar{\Psi}_{S}$,
$\left\langle \overline{\Psi}_{S}\Psi_{D}\right\rangle _{0}\neq0$.
Gap equations are derived for the condensates and the condition of
symmetry breaking is determined. The new fermions get contributions
to their masses from the condensates. The vacuum solution of the model
has a nontrivial weak $SU_{L}(2)$ quantum number and it spontaneously
breakes the electroweak symmetry in a dynamical way. This symmetry
breaking scheme was already utilized in our earlier works \cite{vcm2,vcm3}.
The nontrivial condensate further generates mixing between the neutral
component of the doublet and the singlet. The $\Psi_{D}$ doublet
has a standard kinetic terms with the usual covariant derivative and
after the mixing the weak gauge bosons ($W^{\pm},$ $Z$) get their
masses from the symmetry breaking condensate. The proposed model contains
three new particles, two neutral and a charged fermions. The solution
of the gap equations shows that the mass of the charged fermion is
between the two neutral ones. The lighter neutral particle is an ideal
dark matter candidate. The most important constraints on the parameters
of the model are coming from the solution of the gap equation and
the requirement of perturbative unitarity in two particle elastic
scattering processes. Generally the new charged fermion tends to be
nearly degenerate with the heavier neutral one. Perturbative unitarity
sets an upper bound on the lighter neutral fermion depending on the
range of validity of the model (the cutoff), it is $M_{1}\leq230$
GeV for $\Lambda=3$ TeV.

Any beyond the Standard Model physics must face the tremendous success
of the Standard Model in high energy experiments, it must have evaded
direct detection and fulfill the electroweak precision tests. LEP1
and LEP2 mesurements have set a direct lower bound \cite{pdg} for
a heavy charged strongly not interacting fermion (lepton) $M_{+}>100$
GeV and without assumptions $M_{0}>45$ GeV for neutral one. Oblique
radiative correction which proved to be fatal in case of the original
technicolor models are nearly harmless. The starting vector-like doublet
and singlet gives no contribution to the Peskin-Takeuchi $S$ and
$T$ oblique parameters \cite{stu} and the deviations are always
proportional to the mixing among the new neutral fermions. Small enough
but nonvanishing mixing will break the electroweak symmetry but gives
small $S$ and $T$ . Finally the symmetry breaking solutions of the
gap equations are so specially constrained that lead to a miniscule
$S$ and $T$ parameters.

The rest of the chapter is organized as follows. In section 2 we present
the proposed dynamical symmetry breaking model, then the gap equations
are derived and solved, the solutions are further constrained by perturbative
unitarity in section 4. In section 5 the interactions relevant in
phenomenology and direct constraints from the LEP experiment are calculated.
In section 6 we calculate the oblique electroweak parameters and section
7 contains the numerical results and figures. The cross sections for
the LHC and the next generation of linear colliders are presented
before the conclusion, and one appendix flashes a new regularization
method developed and used by us during this work.

\section{The Fermion Condensate Model}

Recently self-interacting vector-like fermions were introduced \cite{fcm}
in the Standard Model instead of an elementary standard scalar Higgs.
The new colourless Dirac fermions are an extra neutral weak $SU(2)$
singlet $(T=Y=0)$ and a doublet \begin{equation}
\Psi_{S},\qquad\Psi_{D}=\left(\begin{array}{c}
\Psi_{D}^{+}\\
\Psi_{D}^{0}\end{array}\right),\label{eq:d1}\end{equation}
with hypercharge 1. Similar fermions are often dubbed leptons, because
they do not participate in strong interactions, and widely studied
in the literature as we discussed in the introduction. A model with
similar fermion content were studied by Maekawa \cite{maek,maek2}.
There is a new $Z_{2}$ symmetry acting only on the new fermions,
which protects them from mixings with the standard model quarks and
leptons, the new fermions may interact only in pairs. The lightest
new fermion is stable providing an ideal weakly interacting dark matter
candidate.

The new Lagrangian with gauge invariant kinetic terms and invariant
4-fermion interactions of the new fermions is $L_{\Psi}$,\begin{eqnarray}
L_{\Psi} & = & \phantom{+}i\overline{\Psi}_{D}D_{\mu}\gamma^{\mu}\Psi_{D}+i\overline{\Psi}_{S}\partial_{\mu}\gamma^{\mu}\Psi_{S}-m_{0D}\overline{\Psi}_{D}\Psi_{D}-m_{0S}\overline{\Psi}_{S}\Psi_{S}+\nonumber \\
 &  & +\lambda_{1}\left(\overline{\Psi}_{D}\Psi_{D}\right)^{2}+\lambda_{2}\left(\overline{\Psi}_{S}\Psi_{S}\right)^{2}+2\lambda_{3}\left(\overline{\Psi}_{D}\Psi_{D}\right)\left(\overline{\Psi}_{S}\Psi_{S}\right),\label{eq:4fermion}\end{eqnarray}
$m_{0D},m_{0S}$ are bare masses and $D_{\mu}$ is the covariant derivative
\begin{equation}
D_{\mu}=\partial_{\mu}-i\frac{g}{2}\underline{\tau}\,\underline{A}_{\mu}-i\frac{g'}{2}B_{\mu},\label{eq:covariantd}\end{equation}
where $\underline{A}_{\mu,}B_{\mu}$ and $g,\; g'$ are the usual
weak gauge boson fields and couplings, respectively. The left handed
and the right handed fermions are assumed to be gauged under the same
gauge $SU_{L}(2)$ group. Additional four-fermion couplings are possible
but the extra term will not fundamentally change the symmetry breaking
and mass generation. We will show in what follows that for couplings
$\lambda_{i}$ exceeding the critical value the four-fermion interactions
of (\ref{eq:4fermion}) generate condensates 

\begin{eqnarray}
\left\langle \overline{\Psi}_{D\alpha}^{0}\Psi_{D\beta}^{0}\right\rangle _{0} & = & a_{1}\delta_{\alpha\beta},\label{eq:condD}\\
\left\langle \overline{\Psi}_{D\alpha}^{+}\Psi_{D\beta}^{+}\right\rangle _{0} & = & a_{+}\delta_{\alpha\beta},\label{eq:condP}\\
\left\langle \overline{\Psi}_{S\alpha}\Psi_{S\beta}\right\rangle _{0} & = & a_{2}\delta_{\alpha\beta},\label{eq:condS}\\
\left\langle \overline{\Psi}_{S}\Psi_{D}\right\rangle _{0}=\left\langle \left(\begin{array}{c}
\overline{\Psi}_{S}\Psi_{D}^{+}\\
\overline{\Psi}_{S}\Psi_{D}^{0}\end{array}\right)\right\rangle _{0} & \neq & 0.\label{eq:conddoublet}\end{eqnarray}
The formation of the charged condensate (\ref{eq:condP}) first appeared
in \cite{fcmgap} and is more general then the condensates in \cite{fcm}.
The non-diagonal condensate in (\ref{eq:conddoublet}) spontaneously
breaks the group $SU_{L}(2)\times U_{Y}(1)$ to $U_{em}(1)$ of electromagnetism.
With the gauge transformations of $\Psi_{D}$ the condensate (\ref{eq:conddoublet})
can always be transformed into a real lower component,

\begin{equation}
\left\langle \overline{\Psi}_{S\alpha}\Psi_{D\beta}^{0}\right\rangle _{0}=a_{3}\delta_{\alpha\beta},\quad\left\langle \overline{\Psi}{}_{S\alpha}\Psi_{D\beta}^{+}\right\rangle _{0}=0,\label{eq:mixed cond}\end{equation}
where $a_{3}$ is real. The composite operator $\overline{\Psi}_{S}\Psi_{D}$
resembles the standard scalar doublet. 

Assuming invariant four-fermion interactions for the new and known
fermions,\begin{equation}
L_{f}=g_{f}\left(\overline{\Psi}_{L}^{f}\Psi_{R}^{f}\right)\left(\overline{\Psi}_{S}\Psi_{D}\right)+g_{f}\left(\overline{\Psi}_{R}^{f}\Psi_{L}^{f}\right)\left(\overline{\Psi}_{D}\Psi_{S}\right),\label{eq:Yukawa}\end{equation}
the condensate (\ref{eq:mixed cond}) generates masses to the standard
femions. In the linearized, mean field approximation the electron
mass, for example, is\begin{equation}
m_{e}=-4g_{e}a_{3}.\label{eq:melectron}\end{equation}
Up type quark masses can be generated via the charge conjugate field
$\widetilde{\Psi}_{D}=i\tau_{2}\left(\Psi_{D}\right)^{\dagger}.$
Introducing nondiagonal quark bilinears, the Kobayashi-Maskawa mechanism
emerges. As in the Standard Model, from (\ref{eq:melectron}) we see
that for two particles $m_{i}/m_{j}=g_{i}/g_{j}$, the masses are
proportional to the unconstrained generalized Yukawa coefficients.

The masses of the weak gauge bosons arise from the effective interactions
of the auxiliary composite $Y=1$ scalar doublet,\begin{equation}
\Phi=\left(\begin{array}{c}
\Phi^{+}\\
\Phi^{0}\end{array}\right)=\overline{\Psi}_{S}\Psi_{D}.\label{eq:scalardef}\end{equation}
$\Phi$ develops a gauge invariant kinetic term in the low energy
effective description 

\begin{equation}
L_{H}=h\left(D_{\mu}\Phi\right)^{\dagger}\left(D^{\mu}\Phi\right),\label{eq:L phi}\end{equation}
 where $D_{\mu}$ is the usual covariant derivative (\ref{eq:covariantd}).

The coupling constant $h$ sets the dimension of $L_{H}$, $[h]=-4$
in mass dimension, we assume $h>0$. (\ref{eq:L phi}) is a non-renormalizable
Lagrangian and it provides the weak gauge boson masses and some of
the interactions of the new fermions with the standard gauge bosons. 

The terms with $\Phi^{0}$ in $L_{H}$ can be written as\begin{eqnarray}
h^{-1}L_{H} & =\phantom{+} & \frac{g^{2}}{2}W_{\mu}^{-}W^{+\mu}\Phi^{0\dagger}\Phi^{0}+\frac{g^{2}}{4\cdot\cos^{2}\theta_{W}}Z_{\mu}Z^{\mu}\Phi^{0\dagger}\Phi^{0}+\label{eq:hLH}\\
 &  & +\left[\partial^{\mu}\Phi^{0\dagger}\partial_{\mu}\Phi^{0}-\frac{i}{2}\frac{g}{\cos\theta_{W}}\left(\partial^{\mu}\Phi^{0\dagger}\right)\Phi^{0}Z_{\mu}+\frac{i}{2}\frac{g}{\cos\theta_{W}}\Phi^{0\dagger}Z_{\mu}\left(\partial^{\mu}\Phi^{0}\right)\right]\nonumber \end{eqnarray}
in terms of the standard vector boson fields.

In the linearized approximation in (\ref{eq:hLH}) we put

\begin{equation}
h\,\Phi^{0\dagger}\Phi^{0}\rightarrow h\left\langle \Phi^{0\dagger}\Phi^{0}\right\rangle _{0}=h\left(16a_{3}^{2}-4a_{1}a_{2}\right)=\frac{v^{2}}{2},\label{eq:phi vev}\end{equation}
leading to the standard masses\begin{equation}
m_{W}=\frac{gv}{2},\qquad m_{Z}=\frac{gv}{2\cos\theta_{W}}.\label{eq:standardm}\end{equation}
$v^{2}$ is, as usual, $\left(\sqrt{2}G_{F}\right)^{-1}$, $v=254$
GeV . The tree masses naturally fulfill the important relation $\rho_{\mathrm{tree}}=1$.
This relation is the direct consequence of the extra global (custodial)
$SU(2)$ symmetry \cite{cust} of the Lagrangian (\ref{eq:L phi})
and of the vacuum expectation value of the composite scalar field.
The complete symmetry breaking sector, the Lagrangian (\ref{eq:4fermion})
does not show this extra global symmetry, because there are mass-like
terms breaking the symmetry of global chiral rotations. However, this
symmetry breaking does not influence the $W^{\pm}$, $Z$ mass ratio.
The idea is that there is a compositness scale at the order of the
cutoff $\Lambda$, where the vacuum expectation values of the new
fermions and composite field $\Phi$ is formed, which decouples from
the original fermions at lower energies. This way the composite scalar
field $\Phi$ can have separate global custodial symmetry and the
new fermions can only influence the $\varrho$ parameter via suppressed
loop corrections.

\section{Gap equations}

Once the condensates (\ref{eq:condD}-\ref{eq:conddoublet}) are formed,
dynamical mass terms are generated in the Lagrangian (\ref{eq:4fermion})
beside the bare mass terms. 

\begin{equation}
L_{\psi}\rightarrow L_{\Psi}^{\mathrm{lin}}=-m_{+}\overline{\Psi_{D}^{+}}\Psi_{D}^{+}-m_{1}\overline{\Psi_{D}^{0}}\Psi_{D}^{0}-m_{2}\overline{\Psi}_{S}\Psi_{S}-m_{3}\left(\overline{\Psi^{0}}_{D}\Psi_{S}+\overline{\Psi}_{S}\Psi_{D}^{0}\right),\label{eq:fermion mass}\end{equation}
with

\begin{eqnarray}
m_{+} & = & m_{0D}-6\lambda_{1}a_{+}-8\left(\lambda_{1}a_{1}+\lambda_{3}a_{2}\right)=m_{1}+2\lambda_{1}\left(a_{+}-a_{1}\right)\label{eq:mtablp}\\
m_{1} & = & m_{0D}-6\lambda_{1}a_{1}-8\left(\lambda_{1}a_{+}+\lambda_{3}a_{2}\right),\label{eq:mtable}\\
m_{2} & = & m_{0S}-6\lambda_{2}a_{2}-8\lambda_{3}\left(a_{1}+a_{+}\right),\label{eq:mtabl2}\\
m_{3} & = & 2\lambda_{3}a_{3}.\label{eq:mtabl3}\end{eqnarray}
\begin{figure}[h]
\begin{centering}
\includegraphics[scale=0.45]{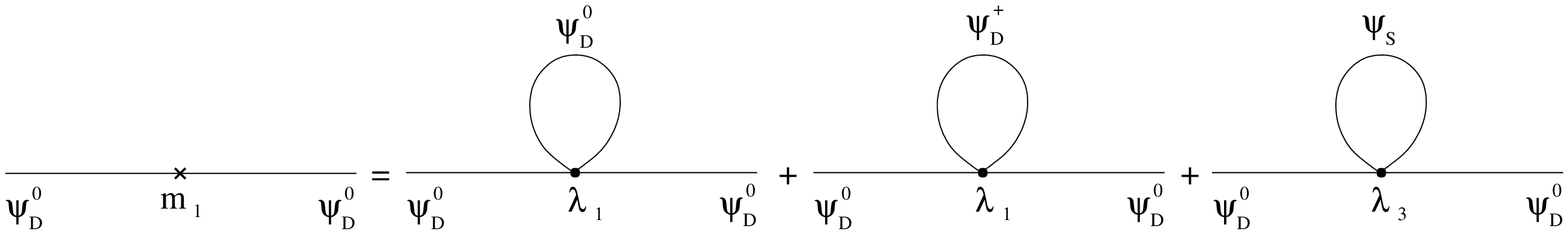}
\par\end{centering}

\begin{centering}
Figure 1. Feynman graphs for the gap equation (\ref{eq:mtable}).
Similar graphs corresponding to (\ref{eq:mtablp},\ref{eq:mtabl2})
with exchanged legs and lines.
\par\end{centering}

\vspace{0.3cm}

\begin{centering}
\includegraphics[scale=0.24]{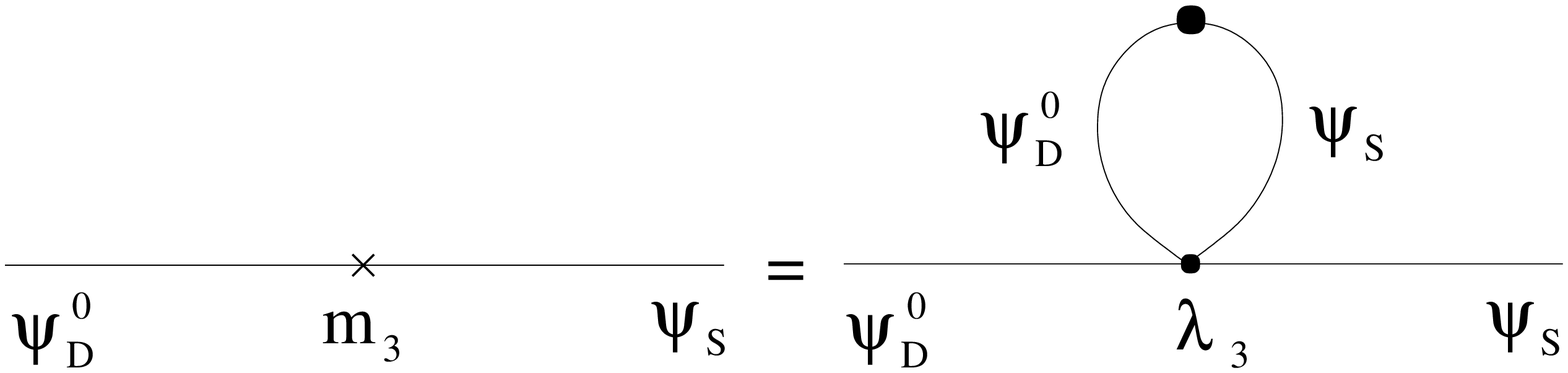}
\par\end{centering}

\begin{centering}
Figure 2. Feynman graphs for the gap equation (\ref{eq:mtabl3}).
\par\end{centering}

\vspace{0.3cm}

\end{figure}
 If $m_{3}=0$ ($\lambda_{3}=0$ or $a_{3}=0$) then (\ref{eq:fermion mass})
is diagonal, the original gauge eigenstates are the physical fields,
the electroweak symmetry is not broken, $\lambda_{3}a_{3}$, the non-diagonal
condensate triggers the mixing and symmetry breaking. If $m_{3}\neq0$
(\ref{eq:fermion mass}) is diagonalized via unitary transformation
to get physical mass eigenstates 

\begin{eqnarray}
\Psi_{1} & = & \phantom{-}c\,\Psi_{D}^{0}+s\,\Psi_{S},\nonumber \\
\Psi_{2} & = & -s\,\Psi_{D}^{0}+c\,\Psi_{S},\label{eq:fermion mixing}\end{eqnarray}
where $c=\cos\phi$ and $s=\sin\phi$, $\phi$ is the mixing angle.
As $\Psi_{S}$ is real only the real components of $\Psi_{D}^{0}$
take part in the mixing. The masses of the physical fermions $\Psi_{1},\:\Psi_{2}$
are

\begin{equation}
2M_{1,2}=m_{1}+m_{2}\pm\frac{m_{1}-m_{2}}{\cos2\phi}.\label{eq:mphys}\end{equation}
The mixing angle is defined by \begin{equation}
2m_{3}=(m_{1}-m_{2})\tan2\phi.\label{eq:def phi}\end{equation}
Again we see, once $m_{3}=0$ the mixing angle vanishes (for $m_{1}\neq m_{2}$),
$M_{1}=m_{1}$ and $M_{2}=m_{2}$. The physical masses wil be equal
($M_{1}=M_{2})$ only if $m_{1}=m_{2}$, the original neutral fermions
are degenerate in mass and then the mixing angle is meaningless from
the point of view of mass matrix diagonalization.

It follows that the physical eigenstates themselves form condensates
since\begin{eqnarray}
c^{2}\left\langle \overline{\Psi}_{1\alpha}\Psi_{1\beta}\right\rangle _{0}+s^{2}\left\langle \overline{\Psi}_{2\alpha}\Psi_{2\beta}\right\rangle _{0} & = & a_{1}\delta_{\alpha\beta},\nonumber \\
s^{2}\left\langle \overline{\Psi}_{1\alpha}\Psi_{1\beta}\right\rangle _{0}+c^{2}\left\langle \overline{\Psi}_{2\alpha}\Psi_{2\beta}\right\rangle _{0} & = & a_{2}\delta_{\alpha\beta},\label{eq:condphys}\\
cs\left\langle \overline{\Psi}_{1\alpha}\Psi_{1\beta}\right\rangle _{0}-cs\left\langle \overline{\Psi}_{2\alpha}\Psi_{2\beta}\right\rangle _{0} & = & a_{3}\delta_{\alpha\beta}.\nonumber \end{eqnarray}
There is no non-diagonal condesate as $\Psi_{1},\:\Psi_{2}$ are independent.
Combining the equations of (\ref{eq:condphys}) one finds

\begin{equation}
a_{3}=\frac{1}{2}\tan2\phi\left(a_{1}-a_{2}\right).\label{eq:a3rel}\end{equation}
For $a_{1}=a_{2}$, $a_{3}\neq0$ is not possible for $\cos2\phi\neq0$.
As is seen, (\ref{eq:a3rel}) is equivalent to $\left\langle \overline{\Psi}_{1\alpha}\Psi_{2\beta}\right\rangle _{0}=0.$
Comparing (\ref{eq:a3rel}) to (\ref{eq:def phi}) yields \begin{equation}
m_{1}-m_{2}=2\lambda_{3}\left(a_{1}-a_{2}\right).\label{eq:m1m2rel}\end{equation}
Using the equations (\ref{eq:mtablp}-\ref{eq:mtabl3}) we are lead
to a consistency conditions\begin{equation}
\left(\lambda_{3}-\lambda_{1}\right)\left(a_{1}+\frac{4}{3}a_{+}\right)=\left(\lambda_{3}-\lambda_{2}\right)a_{2},\label{eq:cons}\end{equation}
$\lambda_{1}\neq\lambda_{2}$ goes with $a_{1}+\frac{4}{3}a_{+}\neq a_{2}$.

The equations (\ref{eq:mtablp}-\ref{eq:mtabl3}) can be formulated
as gap equations \cite{njl} in terms of the physical fields expressing
both the masses and the condensates with $\Psi_{1}$, $\Psi_{2}$
and $\Psi_{+}\equiv\Psi_{D}^{+}$. Assuming vanishing original masses,
$m_{0S}=0$, $m_{0D}=0$, the complete set of gap equations are\begin{eqnarray}
c\cdot s\left(M_{1}-M_{2}\right) & = & 2\lambda_{3}\; c\cdot s\left(I_{1}-I_{2}\right),\label{eq:gap3}\\
c^{2}M_{1}+s^{2}M_{2} & = & -\lambda_{1}\left(6\left(c^{2}I_{1}+s^{2}I_{2}\right)+8I_{+}\right)-8\lambda_{3}\left(s^{2}I_{1}+c^{2}I_{2}\right),\label{eq:gap1}\\
s^{2}M_{1}+c^{2}M_{2} & = & -6\lambda_{2}\left(s^{2}I_{1}+c^{2}I_{2}\right)-8\lambda_{3}\left(c^{2}I_{1}+s^{2}I_{2}+I_{+}\right),\label{eq:gap2}\\
M_{+} & = & -\lambda_{1}\left(8\left(c^{2}I_{1}+s^{2}I_{2}\right)+6I_{+}\right)-8\lambda_{3}\left(s^{2}I_{1}+c^{2}I_{2}\right).\label{eq:gap+}\end{eqnarray}

The main task of the present work is to explore the structure of the
gap equations. There are four algebraic equations for four variables
$M_{1},\; M_{2},\; M_{+}$, $c^{2}=\cos^{2}\phi$. As in almost all
approximation $I_{i}\sim M_{i}$, (\ref{eq:gap3}-\ref{eq:gap+})
show gap equation characteristics, $M_{i}=0$ is always a symmetric
solution, which is stable for small $|\lambda_{i}|$. Increasing $|\lambda_{i}|$
also an energetically favoured \cite{klev} massive solution emerges
as in the original Nambu Jona-Lasinio model. Now we explore the parameter
space $\lambda_{i}$ to find acceptable phyical masses.

Let the condensates be approximated by free field propagators\begin{equation}
\left\langle \overline{\Psi}_{i\alpha}\Psi_{i\beta}\right\rangle =\frac{\delta_{\alpha\beta}}{4}I_{i}=-\frac{\delta_{\alpha\beta}}{8\pi^{2}}M_{i}\left(\Lambda^{2}-M_{i}^{2}\ln\left(1+\frac{\Lambda^{2}}{M_{i}^{2}}\right)\right),\quad i=1,2,+,\label{eq:free1}\end{equation}
where $M_{+}=m_{+}$. Here $\Lambda$ is a four-dimensional physical
cutoff, it sets the scale of the new physics responsible for the non-renormalizable
operators. From the point ov view of symmetry breaking, the $\Lambda$
cutoff can be chosen arbitrary large (below the GUT or Planck scale),
but higher $\Lambda$ implies stronger fine tuning of $\lambda_{3}$,
see (32), to keep the new fermion masses in the electroweak range.
To avoid fine tuning and allow reasonable fermion masses $\Lambda$
is expected to be a few TeV, typically around 3 TeV \cite{fcm}.

For the electroweak symmetry breaking the most important equation
is (\ref{eq:gap3}), it triggers mixing between the different representations
of the weak gauge group. Applying (\ref{eq:free1}) it reads

\begin{equation}
0=\left(M_{1}-M_{2}\right)c\cdot s\left(\frac{1}{\lambda_{3}}+\frac{\Lambda^{2}}{\pi^{2}}-\frac{M_{1}^{3}\ln\left(1+\frac{\Lambda^{2}}{M_{1}^{2}}\right)-M_{2}^{3}\ln\left(1+\frac{\Lambda^{2}}{M_{2}^{2}}\right)}{M_{1}-M_{2}}\right).\label{eq:gap3an}\end{equation}
(\ref{eq:gap3an}) always has a symmetric solution $\left(M_{1}-M_{2}\right)c\cdot s=0$,
implying $\sin2\phi=0$ for $M_{1}\neq M_{2}$, there is essentialy
no mixing. $M_{1}=M_{2}$ is discussed after (\ref{eq:def phi}).
If $|\lambda_{3}|$ is greater than a critical value $\left|\lambda_{3}^{c}\right|=\frac{\pi^{2}}{\Lambda^{2}}$
there also exists a symmetry breaking solution ($M_{1}\neq M_{2})$,
which always has lower energy if the massive solution exists \cite{klev}.
Equation (\ref{eq:gap3an}) has a solution with moderate masses ($M_{1,2}<0.7\Lambda$
) if $\lambda_{3}$ is negative. In the small mass limit the parantheses
in (\ref{eq:gap3an}) simplifies to $\frac{1}{\lambda_{3}}+\frac{\Lambda^{2}}{\pi^{2}}-\left(M_{1}^{2}+M_{1}M_{2}+M_{2}^{2}\right)\left(\ln\left(\Lambda^{2}\right)-\ln\left(\tilde{M^{2}}\right)\right)$
where $\tilde{M}\simeq max(M_{1},M_{2})$. If $|\lambda_{3}|$ is
slightly larger than its critical value, then we generally get small
masses compared to $\Lambda$, $M_{1}^{2}+M_{1}M_{2}+M_{2}^{2}\ll\Lambda^{2}$.
The critical coupling agrees with the original Nambu-Jona Lasinio
value, only a factor of two coming from the definition in the Lagrangian
(\ref{eq:4fermion}). If $\left|\lambda_{3}\right|<\left|\lambda_{3}^{c}\right|$
then the parantheses does not vanish in (\ref{eq:gap3an}), the condensate
$a_{3}$ is not formed and $\left(M_{1}-M_{2}\right)c\cdot s=0$.
The physical solution is $c\cdot s=0$, there is no meaningful mixing,
$\Psi_{S},\,\Psi_{D}$ are the physical mass eigenstates, and the
electroweak symmetry is not broken.

\begin{figure}
\begin{centering}
\includegraphics[scale=0.75]{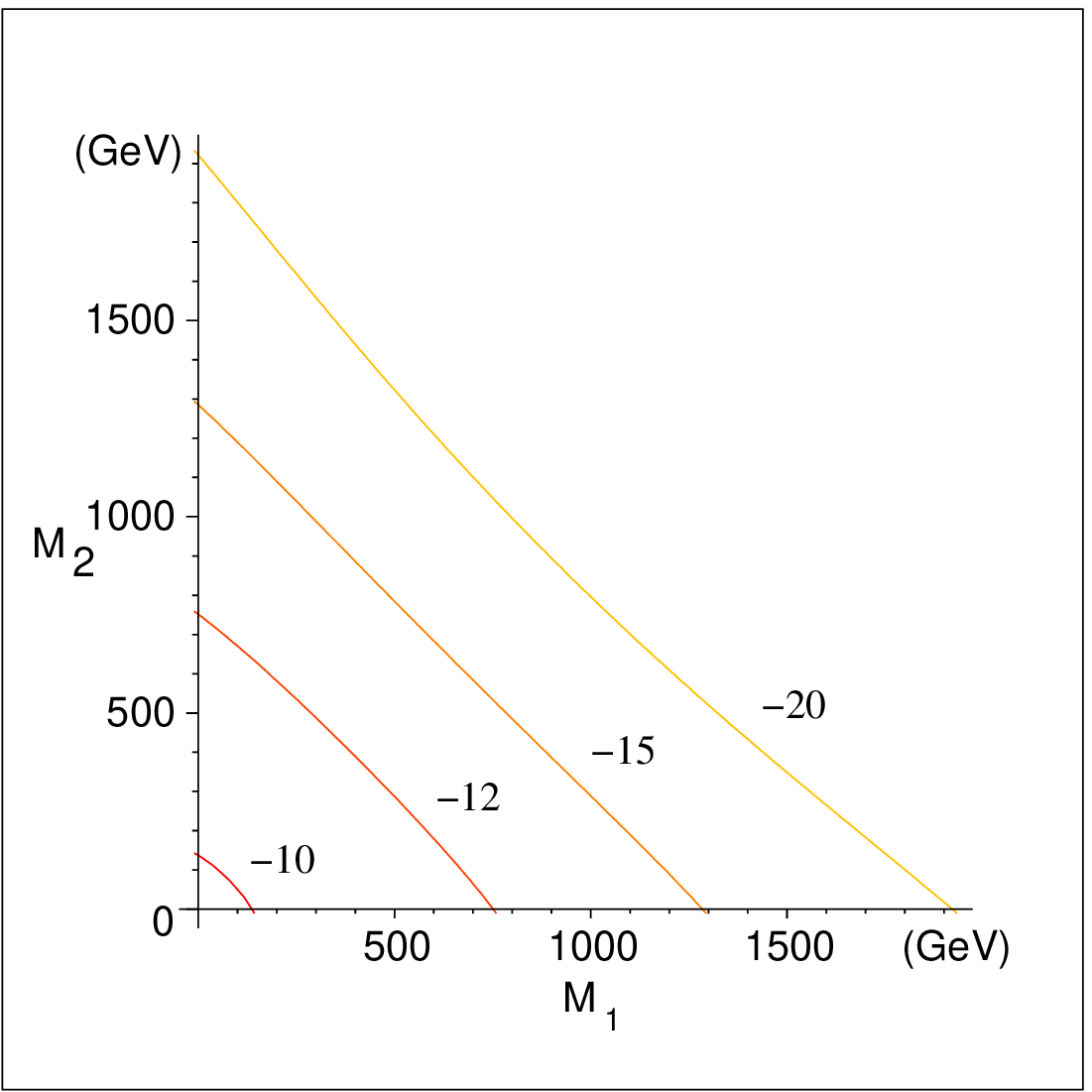}
\par\end{centering}

\centering{}Figure 3. Constant $\lambda_{3}$ contours in the $M_{1}$-$M_{2}$
plane for $\lambda_{3}=\left\{ -10,-12,-15,-20\right\} \cdot1/\Lambda^{2}$,
$\Lambda=3\;$TeV.
\end{figure}

Despite the complicated structure of the non-linear equations (\ref{eq:gap3}-\ref{eq:gap+})
we get a relatively simple gap equation for $\lambda_{1}$, similar
to (\ref{eq:gap3an}), from (\ref{eq:mtablp}) $2\lambda_{1}\left(a_{1}-a_{+}\right)=m_{1}-m_{+}$.
In the physical fields we have\begin{equation}
M_{+}-c^{2}M_{1}-s^{2}M_{2}=2\lambda_{1}\left(I_{+}-c^{2}I_{1}-s^{2}I_{2}\right).\label{eq:l1an}\end{equation}
It includes four unknowns, therefore it cannot be analyzed directly.
We get a useful restriction solving (\ref{eq:gap3}) and (\ref{eq:gap1})
for $\lambda_{1}$ and substituting it to (\ref{eq:l1an}), relating
$M_{1},M_{2},M_{+}$ and $c^{2}$ independently of the $\lambda_{i}$'s.
Requiring that $0\leq c^{2}\leq1$ we get \begin{equation}
M_{1}\leq M_{+}\leq M_{2}.\label{eq:mpconstr}\end{equation}
As a result of the logaritmic terms in $I_{i}$, $M_{+}$ is nonlinear
in $c^{2}$, while $m_{1}=c^{2}M_{1}+s^{2}M_{2}$. We remark that
though (\ref{eq:gap3}) and (\ref{eq:l1an}) are very similar, for
moderate masses $\lambda_{3}$ is always negative, while $\lambda_{1}$
is positive (also $\lambda_{2}>0$). In the $c^{2}=0\;(1)$ limit
$M_{+}=M_{2}\;(M_{1})$ and there are cancellations in (\ref{eq:gap3}-\ref{eq:gap+}).
Turning back to the symmetric solution of (\ref{eq:gap3an}) the relation
(\ref{eq:mpconstr}) gives $M_{+}=M_{1}=M_{2}$ and the rest of the
gap equations set the common mass equal to zero unless the special
relation $6(\lambda_{3}-\lambda_{2})=8(\lambda_{3}-\lambda_{1})$
holds to provide cancellations.

To find the critical value for $\lambda_{1}$ and $\lambda_{2}$ we
considered the limit $M_{+}\rightarrow M_{2}=M$ and $M_{1}\rightarrow0$
then \begin{equation}
\lambda_{1}=\frac{1}{7}\frac{\pi^{2}}{\Lambda^{2}-M^{2}\ln\left(1+\frac{\Lambda^{2}}{M^{2}}\right)},\quad\lambda_{2}=\frac{4}{3}\frac{\pi^{2}}{\Lambda^{2}-M^{2}\ln\left(1+\frac{\Lambda^{2}}{M^{2}}\right)}.\label{eq:l1crit}\end{equation}
We get the same NJL type expression if we take the limit $M_{+}\rightarrow M_{2}=M$
and $M_{1}\rightarrow0$. (\ref{eq:l1crit}) provides massive solutions
if $\lambda_{1}\geq\frac{1}{7}\frac{\pi^{2}}{\Lambda^{2}}$ and $\lambda_{2}\geq\frac{4}{3}\frac{\pi^{2}}{\Lambda^{2}}$
. Numerical scans show that these are the minimal, critical values
for the couplings and can be approximated in special limits. Numerical
solutions are shown in Table 1. for cutoff $\Lambda=3$ TeV. The role
of $M_{1}$ and $M_{2}$ can be exchanged together with $c^{2}\leftrightarrow s^{2}$,
therefore we have chosen $M_{1}<M_{2}$ without the loss of generality.
As the cutoff is not too high, 3 TeV, there is no serious fine tuning
in the $\lambda_{i}$'s to find relatively small masses. 

To understand the signs and roughly the factors in $\lambda_{1,2}^{c}$
consider the limit $M_{1}\simeq M_{2}\simeq M_{+}\simeq M$. If $M\ll\Lambda$
then $\lambda_{3}\simeq\lambda_{3}^{c}=-\frac{\pi^{2}}{\Lambda^{2}}$,
though in the exact limit (\ref{eq:gap3}) becomes singular. We get
from (\ref{eq:gap3}-\ref{eq:gap+}) the relation $14\lambda_{1}=6\lambda_{2}+8\lambda_{3}$
and a single gap equation ( $I=I_{M}$ in (\ref{eq:free1}) )\begin{equation}
M=-\left(14\lambda_{1}+8\lambda_{3}\right)I.\label{eq:smallm}\end{equation}
Small mass solution requires $\tilde{\lambda}=14\lambda_{1}+8\lambda_{3}$
to be close to it's critical value $2\pi^{2}/\Lambda^{2}$ and provides
rough estimates $\lambda_{1}\sim\frac{5}{7}\frac{\pi^{2}}{\Lambda^{2}}$
and also $\lambda_{2}\sim3\frac{\pi^{2}}{\Lambda^{2}}$ to generate
small masses. Numerical solutions also provide general ($M_{+}$ not
close to $M_{1}$ or $M_{2}$) small masses for couplings close to
these values, see Table 1.%
\begin{table}[th]
\begin{centering}
\begin{tabular}{|c|c|c|c|c|c|c|c|}
\hline 
$\lambda_{1}$$\left(\frac{\pi^{2}}{\Lambda^{2}}\right)$ & 0.546 & 0.740 & 0.496 & 0.380 & 0.502 & 0.468 & 0.419\tabularnewline
\hline 
$\lambda_{2}$$\left(\frac{\pi^{2}}{\Lambda^{2}}\right)$ & 2.540 & 3.11 (!) & 2.403 & 2.120 & 2.457 & 2.455 & 2.451\tabularnewline
\hline 
$\lambda_{3}$$\left(\frac{\pi^{2}}{\Lambda^{2}}\right)$ & -1.031 & -1.041 & -1.042 & -1.070 & -1.083 & -1.178 & -1.330\tabularnewline
\hline
\hline 
$M_{1}$ (GeV) & 100 & 148 & 100 & 100 & 150 & 200 & 200\tabularnewline
\hline 
$M_{2}$ (GeV) & 150 & 150 & 200 & 300 & 300 & 500 & 800\tabularnewline
\hline 
$M_{+}$ (GeV) & 149 & 149 & 190 & 290 & 290 & 490 & 790\tabularnewline
\hline
\end{tabular}
\par\end{centering}

\caption{Solutions of the gap equations for the cutoff $\Lambda=$3 TeV, $\lambda_{i}$
are given in units of $\frac{\pi^{2}}{\Lambda^{2}}$. In the second
column $\lambda_{2}$ violates perturbative unitarity.}

\end{table}

In the strongest small mass limit one neglects the logaritmic terms
in the condensates (\ref{eq:free1}), and equations (\ref{eq:gap3}-\ref{eq:gap+})
reduce to a linear homogeneous system of equations \cite{fcmgap}.
Finally we get two relations for the masses, $M_{+}=m_{1}=c^{2}M_{1}+s^{2}M_{2}$
and $\frac{m_{1}}{m_{2}}=\frac{1-6\lambda_{2}\Lambda^{2}/\pi^{2}}{16\lambda_{3}\Lambda^{2}/\pi^{2}}=\frac{8\lambda_{3}\Lambda^{2}/\pi^{2}}{1-14\lambda_{1}\Lambda^{2}/\pi^{2}}$.

The solutions of the gap equations are further constrained by perturbative
unitarity.

\section{Perturbative unitarity}

In this section we apply tree-level partial wave unitarity to two-body
scatterings of the new fermions following the arguments of the pioneering
work by Lee et al. \cite{unit}, where perturbative unitarity has
been employed to constrain the Standard Model Higgs mass. Perturbative
unitarity is a powerful tool, it can be used to build up the bosonic
sector of the Standard Model, moreover it was essential to build higssless
models of electroweak symmetry breaking in extra dimensional field
theories \cite{csaki}. The method was used to constrain the parameters
in the dynamical symmetry breaking vector condensate model in \cite{unitvcm}.

Consider the amplitudes of two particle $\left(\Psi_{D}^{(+)},\Psi_{D}^{(0)}\,\mathrm{or}\;\Psi_{S}\right)$
elastic scattering processes and impose $\left|\Re a_{0}\right|\leq1/2$
for the $J=0$ partial wave amplitudes. The contact graph gives the
dominant contribution, neglecting the fermion masses for the $\Psi_{D}^{(+)}\Psi_{D}^{(-)}$
scattering gives an upper bound on $\lambda_{1}$ coupling, $\left|\lambda_{1}\right|s\leq8\pi$,
where $s$ is the maximal center of mass energy $\left(M_{+}^{2}\ll s\leq\Lambda^{2}\right)$. 

\begin{figure}[h]
\begin{centering}
\includegraphics[scale=0.75]{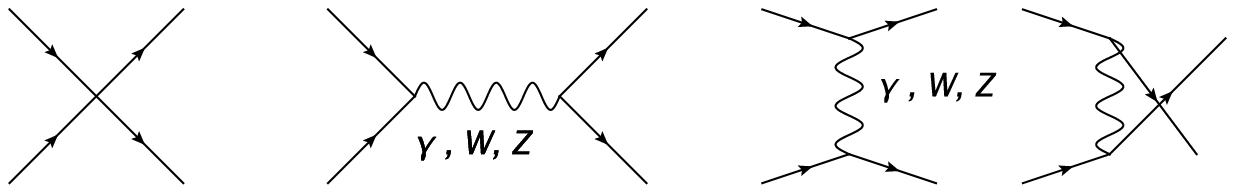}
\par\end{centering}

\centering{}Figure 4. Feynman graphs of 2-particle elastic scattering
\end{figure}

We cannot always use the small mass limit, as the solution of the
gap equations provide higher $\lambda_{i}$'s for significantly higher
masses. Therefore we have calculated different helicity amplitudes
\cite{unitfermion} for non-vanishing masses. For $\Psi_{a}(1)\bar{\Psi}_{a}(2)\rightarrow\Psi_{a}(3)\bar{\Psi}_{a}(4)$,
($a=0,\, s,+$), $M=\lambda_{i}\left[\left(\bar{v}_{2}u_{1}\right)\left(\bar{u}_{3}v_{4}\right)-\left(\bar{u}_{3}u_{1}\right)\left(\bar{v}_{2}v_{4}\right)\right]$,
where $\lambda_{i=1,2,3}$ are the only relevant four-fermion couplings.
We consider $\Psi_{S}=s\Psi_{1}+c\Psi_{2}$ scattering as a linear
combination in the coupled $\Psi_{1},\,\Psi_{2}$ channels to employ
only $\lambda_{2}$ (and simiarly $\Psi_{D}^{(0)}$ to constrain $\lambda_{1}$).
The contributions of the $\gamma,\, Z$ exchange graphs are negligible
$\left({\cal O}\left(g^{2}\right)\ll8\pi\right)$ because of the extra
propagator. There are three different helicity channels, we give the
representative helicity amplitudes, these are maximal for the back
to back scattering ($\theta_{\mathrm{scattering}}^{\left\{ 13\right\} }=\pi$)
\begin{eqnarray}
M\left((+-)\rightarrow(+-)\right) & = & \lambda_{i}\left(s-4M_{i}^{2}\right),\label{eq:hel1}\\
M\left((++)\rightarrow(--)\right) & = & \lambda_{i}s,\label{eq:hel2}\\
M\left((+-)\rightarrow(-+)\right) & = & \lambda_{i}4M_{i}^{2}.\label{eq:hel3}\end{eqnarray}
\begin{figure}[t]
\begin{centering}
\includegraphics[scale=0.5]{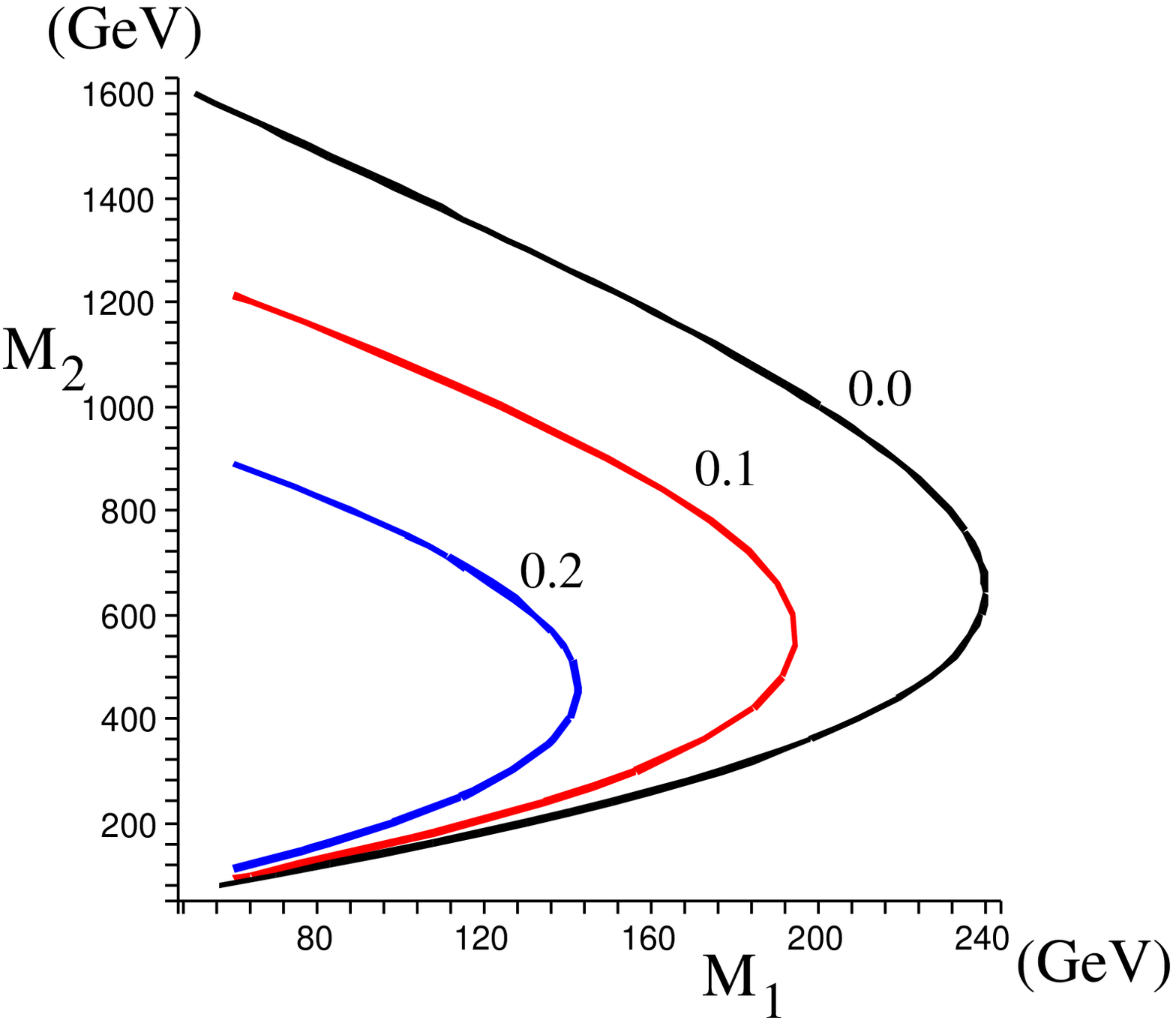} 
\par\end{centering}

Figure 5.\textbf{ }The maximum value of $c{}^{2}=\cos^{2}\Phi$ on
the $M_{1},\, M_{2}$ plane from the gap equation and unitarity. $c^{2}$
can be higher inside the curves.
\end{figure}
For other scattering angles $\left|M\right|$ is smaller than in (\ref{eq:hel2}),
for example the maximum for $\theta=0$ is $\lambda_{i}4M_{i}^{2}$.
The mass dependent unitarity bound agrees with the first estimate\begin{equation}
\lambda_{i}s\leq8\pi,\label{eq:unit}\end{equation}
where i=1,2,3 and $s\leq\Lambda^{2}$ is the center of mass energy.
The unitarity constraints are most stringent for $\lambda_{2}$, even
the equal small mass limit (\ref{eq:smallm}) would set $\lambda_{2}\simeq3\pi^{2}/\Lambda^{2}$
which is above the maximum value allowed by unitarity $8\pi/\Lambda^{2}\simeq2.55\cdot\frac{\pi^{2}}{\Lambda^{2}}$.
As an example we show a non-physical nearly equal mass solution in
the second column of Table 1., which is not allowed by perturbative
unitarity. (\ref{eq:unit}) implies an absolute upper bound on the
smaller neutral mass, $M_{1}<240$ GeV for $\Lambda=3$ TeV. Perturbative
unitarity for $\lambda_{2}$ and the solution of the gap equations
generally push up the charged mass close to $M_{2}$ and sets the
mixing angle $\sin\phi$ close to 1 in (\ref{eq:fermion mixing})
meaning that there is only a small mixing, $\Psi_{2}$ is mostly composed
of $\Psi_{D}^{0}$ and there is only a small mass splitting in the
doublet $\Psi_{D}$ after symmetry breaking. This observation will
be important to estimate electroweak oblique corrections. The allowed
$M_{1},\, M_{2}$ masses and the maximum value of $c^{2}$ is shown
in Figure 5. The maximum value of the cosine of mixing angle is determined
from the condition that $\lambda_{2}$ should stay below the unitarity
bound \eqref{eq:unit}. The charged fermion mass must be relatively
close to the mass of the heavier neutral one. The mixing angle $\phi$
is relatively close to $\cos\phi\sim0$, the mixing is weak, see the
curve on the right in Figure 6. $\Psi_{2}$ is mostly composed of
$\Psi_{D}^{0}$ and there is only a small mass splitting in the doublet
$\Psi_{D}$ after symmetry breaking.

\section{Interactions with $W^{\pm},$$Z$ and constraints from the Z decay}

The collider phenomenology and radiative corrections (see section
6) in the model are coming from the doublet kinetic term in (\ref{eq:4fermion})
taking into account the mixing (\ref{eq:fermion mixing})

\begin{eqnarray}
L^{I} & = & \phantom{+}\overline{\Psi_{D}^{+}}\gamma^{\mu}\Psi_{D}^{+}\left(\frac{g'}{2}B_{\mu}+\frac{g}{2}W_{3\mu}\right)+\nonumber \\
 &  & +\left(c^{2}\overline{\Psi}_{1}\gamma^{\mu}\Psi_{1}+s^{2}\overline{\Psi}_{2}\gamma^{\mu}\Psi_{2}-sc\left(\overline{\Psi}_{1}\gamma^{\mu}\Psi_{2}+\overline{\Psi}_{2}\gamma^{\mu}\Psi_{1}\right)\right)\left(\frac{g'}{2}B_{\mu}-\frac{g}{2}W_{3\mu}\right)+\nonumber \\
 &  & +\left[\frac{g}{\sqrt{2}}W_{\mu}^{+}\left(c\overline{\Psi_{D}^{+}}\gamma^{\mu}\Psi_{1}-s\overline{\Psi_{D}^{+}}\gamma^{\mu}\Psi_{2}\right)+h.c.\right].\label{eq:Llmix}\end{eqnarray}
The interactions between the new and the standard fermions in $L_{f}$
(\ref{eq:Yukawa}) turns out to be very weak. Indeed, from (\ref{eq:melectron})
and (\ref{eq:phi vev}) we have an upper bound for $g_{e}$, $g_{e}\leq\sqrt{2h}\frac{m_{e}}{v}=\sqrt{2h}g_{e}^{SM}$,
which is suppressed by two factors of the scale of new physics compared
to the standard model value $g_{e}^{SM}$.

We will explore the consequences of these interactions in the decay
of the $Z$ boson and in the precision electroweak test of the standard
model in the next section.

The proposed new fermions could not be seen in the high energy experiments
so far, because of their large masses and/or small couplings to ordinary
particles. The mixing in the doublet reduces the coupling to the gauge
bosons, but the new charged fermion is not affected. From the LEP1
and LEP2 measurements there is lower bound for the mass of a heavy
charged lepton, valid here $M_{+}>100$ GeV \cite{pdg}. For the neutral
component of the doublet (without mixing) there are smaller lower
bounds; without further assumptions $M_{2}>45$ GeV. Using the relation
\eqref{eq:mpconstr} $M_{2}$ is at least 100 GeV with or without
mixing. %
\begin{figure}[t]
\begin{centering}
\includegraphics[scale=0.5]{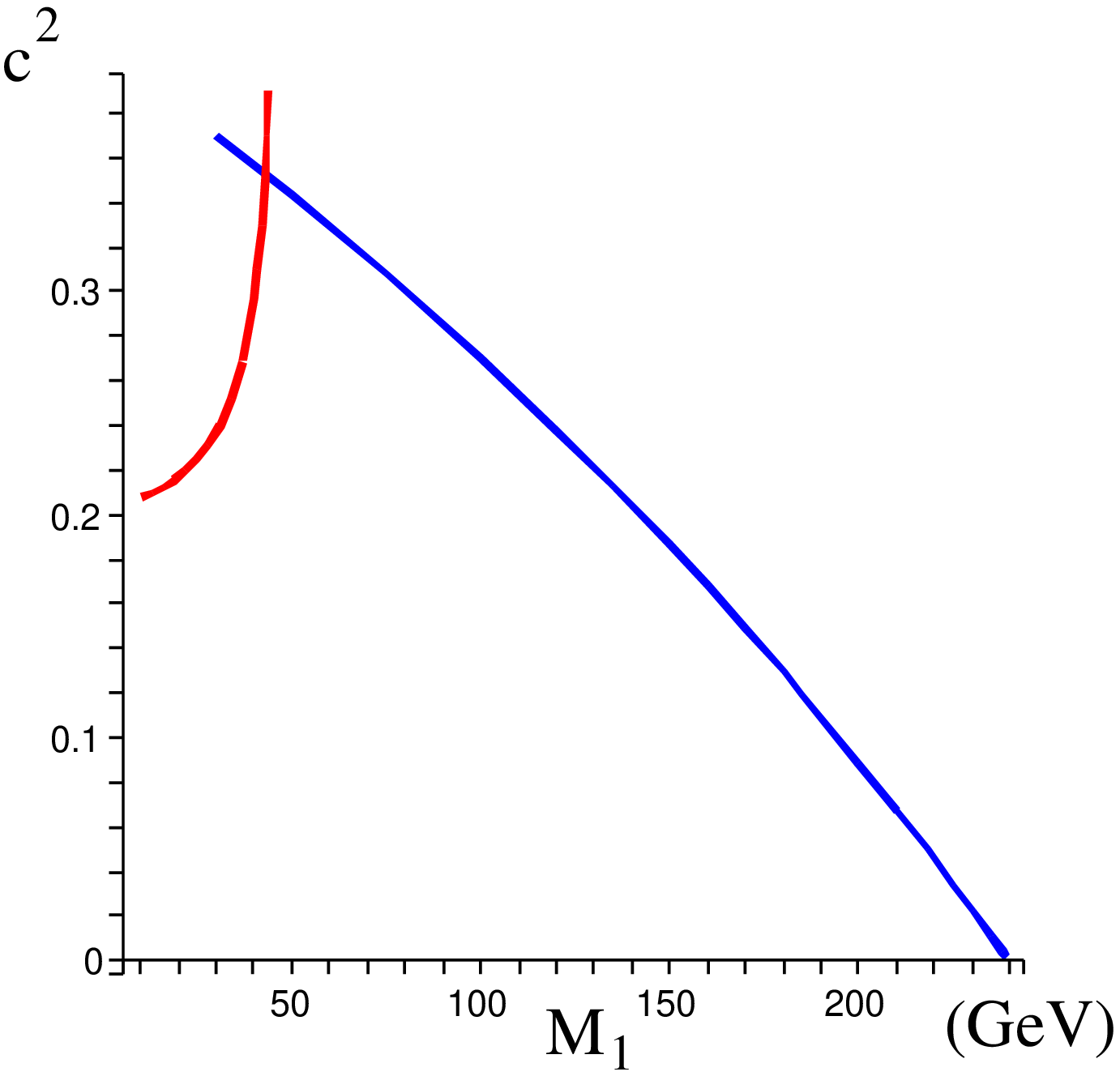}
\par\end{centering}

\centering{}Figure 6. The maximum value of the $c{}^{2}=\cos^{2}\Phi$
vs. the lighter neutral mass $M_{1}$. The right (blue) curve is derived
from the gap equation and unitarity. The upper left (red) curve is
from the width of the Z boson.
\end{figure}
The mixing generates small, but non-vanishing coupling between the
Z boson and the new lighter neutral fermion (e.g. the remnant of the
singlet, it has $c^{2}$ part of a doublet). Therefore if it is light
enough it contributes to the invisible width of the Z boson \begin{equation}
\Gamma(Z\rightarrow\bar{\Psi}_{1}\Psi_{1})=\frac{\sqrt{2}{\mathrm{G}_{F}M_{Z}^{3}}}{6\pi}\left(\frac{c^{4}}{4}\right)\sqrt{1-\frac{4M_{1}^{2}}{M_{Z}^{2}}}.\end{equation}
 The Z width is experimentally known at high precision and the pull
factor is rather small \begin{equation}
\Gamma(Z)=(2.4952\pm0.0023)\hbox{GeV}.\end{equation}
 We estimate the maximum possible room for new physics as 3$\sigma$
in the experimental Z width, $\Gamma_{Z}^{\hbox{new}}<$7 MeV. In
\cite{Pocsik} the minimum value of $\Gamma_{Z}^{\hbox{theory}}$
(at maximum $\sin^{2}\theta_{W}$ and minimum $M_{Z}^{2}$ and $\alpha_{S}$)
was compared to the maximal experimental value, and gave a similar
$3\sigma$ window for new physics. We see that $M_{1}$ masses well
below $M_{Z}/2$ are still allowed for rather small mixing, see the
(red) curve on the left on Figure 6.

\section{Electroweak precision parameters}

The new fermions have direct interactions with the standard fermions
\eqref{eq:Yukawa} and gauge bosons \eqref{eq:Llmix}. The four-fermion
couplings of the new particles to the light fermions are weak; weaker
than the corresponding ones in the Standard Model \cite{fcm}. The
new couplings to the gauge bosons are the gauge couplings suppressed
only by the ${\cal O}(1)$ mixing factors. Therefore the couplings
to the light fermions which participate in the precision experiments,
are suppressed compared to the couplings to the gauge bosons. The
new fermions thus mainly contribute to the gauge boson self energies
in the precision experiments. In most of the solutions of the gap
equation \cite{fcmgap} $M_{+},M_{2}\gg M_{Z}$. Expecting further
$M_{1}>M_{Z}$ we can give a good estimate of the effects of new physics
in terms of the general S, T and U parameters introduced by Peskin
and Takeuchi \cite{stu}. We get a rough estimate of the loop effects
if the mass of the lighter neutral fermion is not far above the $Z$
mass.

The two relevant parameters, $S$ and $T$ defined via the gauge boson
self energies \begin{eqnarray}
\alpha(M_{Z})\, T & = & \frac{\Pi_{WW}^{{\rm new}}(0)}{M_{W}^{2}}-\frac{\Pi_{ZZ}^{{\rm new}}(0)}{M_{Z}^{2}},\\
\frac{\alpha(M_{Z})}{4s_{W}^{2}c_{W}^{2}}\, S & = & \frac{\Pi_{ZZ}^{{\rm new}}(M_{Z}^{2})-\Pi_{ZZ}^{{\rm new}}(0)}{M_{Z}^{2}}-\frac{c_{W}^{2}-s_{W}^{2}}{c_{W}s_{W}}\frac{\Pi_{Z\gamma}^{{\rm new}}(M_{Z}^{2})}{M_{Z}^{2}}-\frac{\Pi_{\gamma\gamma}^{{\rm new}}(M_{Z}^{2})}{M_{Z}^{2}},\label{eq:stuPT}\end{eqnarray}
 where $s_{W}^{2}=\sin^{2}{\theta}_{W}(M_{Z})$ and $c_{W}^{2}=\cos^{2}{\theta}_{W}(M_{Z})$
are $\sin^{2}$ ($\cos^{2}$) of the weak mixing angle. Barbieri et
al. \cite{lep2} revised the definition of the oblique parameters.
The $\Pi$ functions are defined from the transverse gauge boson vacuum
polarization amplitudes expanded around zero $\Pi_{ab}(q^{2})\simeq\Pi_{ab}(0)+q^{2}\Pi'_{ab}(0)+1/2\cdot q^{2}\Pi''_{ab}(0)+...$,
(a,b = 1,3,Y) up to second order. The 12 coefficients define 7 parameter
at the end. The definitions of the old parameters are \begin{eqnarray}
\frac{\alpha(M_{Z})}{4s_{W}^{2}c_{W}^{2}}S & = & \Pi_{3Y}^{\prime\,{\rm new}}(0)\\
\alpha(M_{Z})T & = & \frac{1}{M_{W}^{2}}\left(\Pi_{33}^{{\rm new}}(0)-\Pi_{11}^{{\rm new}}(0)\right),\\
\frac{\alpha(M_{Z})}{4s_{W}^{2}}U & = & \Pi_{33}^{\prime{\rm new}}(0)-\Pi_{11}^{\prime{\rm new}}(0).\label{eq:stujo}\end{eqnarray}
 These parameters (with the extra 4 -$V,\, X,\, Y$ and $W$ ) fall
into three groups according to their symmetry properties \cite{lep2}.
The Peskin-Takeuchi $S$ parameter is custodially symmetric but weak
isospin breaking. The $T$ and $U$ parameters break both the custodial
and the weak isospin symmetry. It is reasonable to expect (and the
actual calculation justifies the assumption) that the parameters with
the same symmetry properties are related to each other. Since $U$
mainly differs from $T$ by an extra derivation of the $\Pi$ functions,
$U\sim\frac{M_{W}^{2}}{M_{{\rm new}}^{2}}T$ is expected, where $M_{{\rm new}}$
is the mass scale of new physics. When there is a gap between $M_{{\rm new}}$
and $M_{W}$ it is reasonable to keep only the lowest derivative terms
with a given symmetry property, $S$ and $T$. If there is no special
fine tuning $U$ is expected to be less important than $T$ and $S$
is kept as the leading effect in its symmetry class.

\begin{figure}
\selectlanguage{magyar}%
\begin{centering}
\inputencoding{latin2}\includegraphics[scale=0.35]{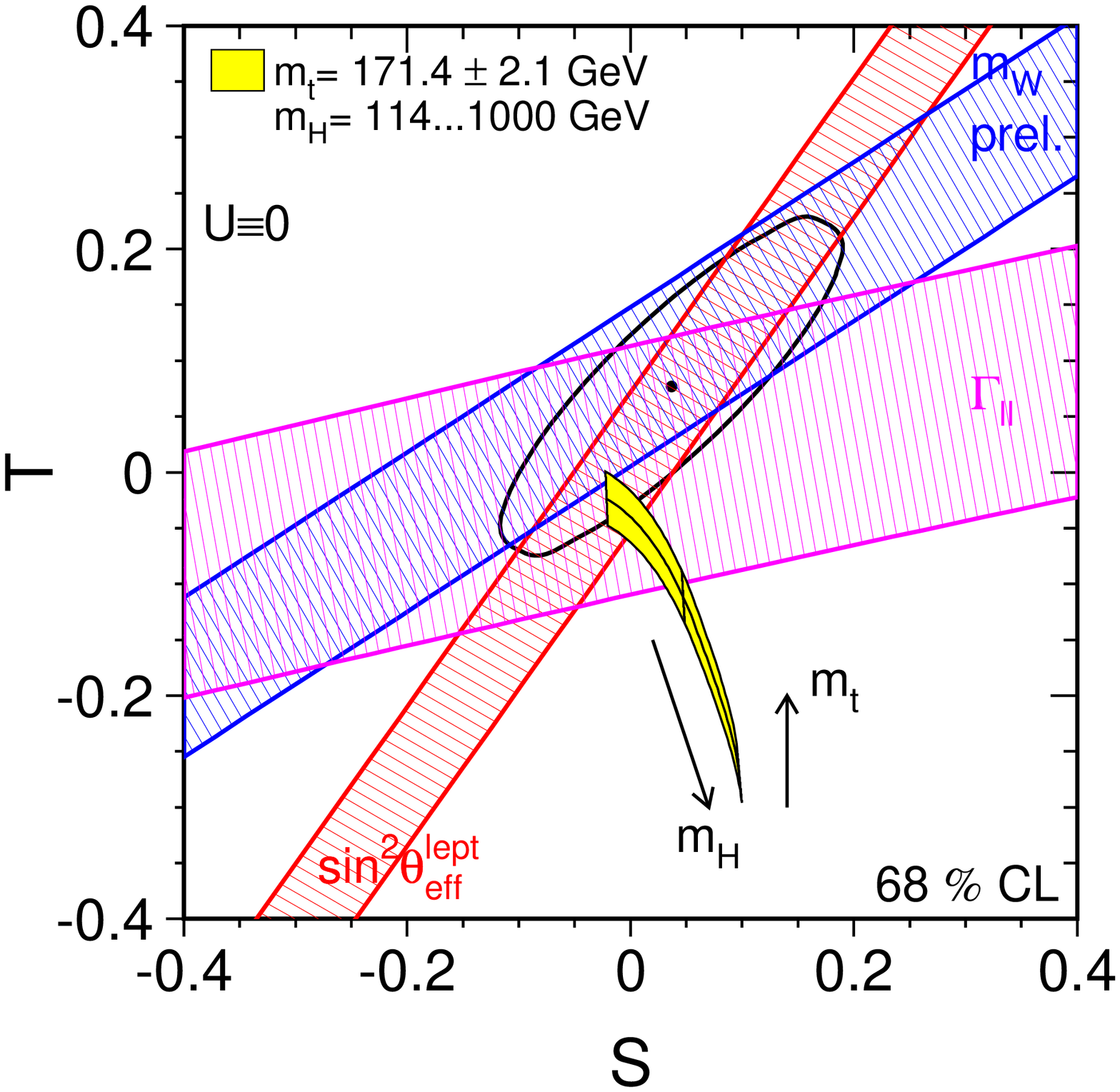}
\par\end{centering}

\selectlanguage{english}%
\centering{}Figure 7. Experimental constraints and Standard Model
predictions for S and T \cite{LEPEWWG}. 
\end{figure}

The experimental data determines $S$, $T$ and $U$ \cite{pdg} \begin{eqnarray}
S & = & -0.10\pm0.10\;(-0.08),\label{Sexp}\\
T & = & -0.08\pm0.11\;(+0.09),\label{Texp}\\
U & = & +0.15\pm0.11\;(+0.01),\label{eq:stuexp}\end{eqnarray}
where the central value assumes $M_{H}=117$ GeV and in parentheses.
The difference is shown for $M_{H}=300$ GeV. The various experimental
constraints and the dependence on the top and Higgs mass can be seen
in Figure 7. In our model the Higgs mass of the fit is understood
as the contribution of a composite Higgs particle with the given mass.

The contributions of the new sector to the gauge boson vacuum polarizations
are fermion loops with generally two non-degenerate masses $m_{a}$
and $m_{b}$ \cite{fcmlambda}. In the low energy effective model
we have preformed the calculation with a 4-dimensional Euclidean momentum
cutoff $\Lambda$. The coupling constants are defined in the usual
manner $L^{I}\sim V_{\mu}\bar{\Psi}\left(g_{V}\gamma^{\mu}+g_{A}\gamma_{5}\gamma^{\mu}\right)\Psi$
\begin{equation}
\Pi(q^{2})=\frac{1}{4\pi^{2}}\left(g_{V}^{2}\,\tilde{\Pi}_{V}+g_{A}^{2}\,\tilde{\Pi}_{A}\right).\label{eq:pivv}\end{equation}
The electroweak parameters depend on the values and derivatives of
the $\Pi$ functions at $q^{2}=0$ \begin{eqnarray}
\tilde{\Pi}_{V}(0) & = & \frac{1}{4}(m_{a}^{2}+m_{b}^{2})-\frac{1}{2}\left(m_{a}-m_{b}\right)^{2}\ln\left(\frac{\Lambda^{2}}{m_{a}m_{b}}\right)-\label{eq:piv0}\\
 &  & -\frac{m_{a}^{4}+m_{b}^{4}-2m_{a}m_{b}\left(m_{a}^{2}+m_{b}^{2}\right)}{4\left(m_{a}^{2}-m_{b}^{2}\right)}\ln\left(\frac{m_{b}^{2}}{m_{a}^{2}}\right).\nonumber \end{eqnarray}
 The first derivative is \begin{eqnarray}
\tilde{\Pi}'_{V}(0) & \!\!=\!\!\! & -\frac{2}{9}-\frac{4m_{a}^{2}m_{b}^{2}-3m_{a}m_{b}\left(m_{a}^{2}+m_{b}^{2}\right)}{6\left(m_{a}^{2}-m_{b}^{2}\right)^{2}}+\frac{1}{3}\ln\left(\frac{\Lambda^{2}}{m_{a}m_{b}}\right)+\label{eq:piv+av0}\\
 &  & +\frac{\left(m_{a}^{2}+m_{b}^{2}\right)\left(m_{a}^{4}-4m_{a}^{2}m_{b}^{2}+m_{b}^{4}\right)+6m_{a}^{3}m_{b}^{3}}{6\left(m_{a}^{2}-m_{b}^{2}\right)^{3}}\ln\left(\frac{m_{b}^{2}}{m_{a}^{2}}\right).\nonumber \end{eqnarray}
 For completeness we give the second derivative, too. It can be used
to calculate further precision parameters \cite{lep2,stu6} e.g. extra
two parameters introduced by Barbieri et al., and it is presented
for extra vector-like fermions in \cite{fcmew}, \begin{eqnarray}
\tilde{\Pi}''_{V}(0) & = & \frac{\left(m_{a}^{2}+m_{b}^{2}\right)\left(m_{a}^{4}-8m_{a}^{2}m_{b}^{2}+m_{b}^{4}\right)}{8\left(m_{a}^{2}-m_{b}^{2}\right)^{4}}+\frac{m_{a}m_{b}\left(m_{a}^{4}+10m_{a}^{2}m_{b}^{2}+m_{b}^{4}\right)}{6\left(m_{a}^{2}-m_{b}^{2}\right)^{4}}-\\
 &  & -\frac{m_{a}^{3}m_{b}^{3}\left(3m_{a}m_{b}-2m_{a}^{2}-2m_{b}^{2}\right)}{2\left(m_{a}^{2}-m_{b}^{2}\right)^{5}}\ln\left(\frac{m_{b}^{2}}{m_{a}^{2}}\right).\label{eq:piv+avv0}\end{eqnarray}
 We get the functions for axial vector coupling by flipping exactly
one of the masses in the previous results ($m_{a}\rightarrow m_{a}$
and $m_{b}\rightarrow-m_{b}$). The method of our calculation has
nice properties: it has no quadratic divergence as expected; it fulfills
gauge invariance in two aspects, $\Pi_{V}(m_{a},m_{a},0)=0$ and the
complete $\Pi$ function is transverse, the coefficients of the $g_{\mu\nu}$
and $-p_{\mu}p_{\nu}/p^{2}$ parts are equal.

The values of the vacuum polarizations for identical masses $(m_{b}=m_{a})$
are smooth limits and agree with direct calculation. \begin{eqnarray}
\tilde{\Pi}_{V\!}(0)=0,\quad\tilde{\Pi}'_{V\!}(0)=-\frac{1}{3}+\frac{1}{3}\ln\left(\frac{\Lambda^{2}}{m_{a}^{2}}\right),\quad\tilde{\Pi}''_{V\!}(0)=\frac{2}{15}\frac{1}{m_{a}^{2}}.\end{eqnarray}
 The ${S}$ parameter is then given by (for the sake of simplicity
the index $V$ is omitted) \begin{equation}
{S}=\frac{1}{\pi}\left(+\tilde{\Pi}'(M_{+},M_{+},0)-c^{4}\tilde{\Pi}'(M_{1},M_{1},0)-s^{4}\tilde{\Pi}'(M_{2},M_{2},0)-2s^{2}c^{2}\tilde{\Pi}'(M_{2},M_{1},0)\right).\label{eq:spar}\end{equation}
 The first three terms cancel the divergent contribution of the last
one.

The ${T}$ parameter related to $\Delta\rho$ is \begin{eqnarray}
{T} & = & \frac{1}{4\pi s_{W}^{2}M_{W}^{2}}\left[+\tilde{\Pi}(M_{+},M_{+},0)+c^{4}\tilde{\Pi}(M_{1},M_{1},0)+s^{4}\tilde{\Pi}(M_{2},M_{2},0)+\right.\nonumber \\
 &  & \left.+2s^{2}c^{2}\tilde{\Pi}(M_{2},M_{1},0)-2c^{2}\tilde{\Pi}(M_{+},M_{1},0)-2s^{2}\tilde{\Pi}(M_{+},M_{2},0)\right].\label{eq:tpar}\end{eqnarray}
 For completeness we give the $U$ parameter in the model \begin{eqnarray}
{U} & =- & \frac{1}{\pi}\left[+\tilde{\Pi}'(M_{+},M_{+},0)+c^{4}\tilde{\Pi}'(M_{1},M_{1},0)+s^{4}\tilde{\Pi}'(M_{2},M_{2},0)+\right.\nonumber \\
 &  & \left.+2s^{2}c^{2}\tilde{\Pi}'(M_{2},M_{1},0)-2c^{2}\tilde{\Pi}'(M_{+},M_{1},0)-2s^{2}\tilde{\Pi}'(M_{+},M_{2},0)\right].\label{eq:upar}\end{eqnarray}

The gauge boson self-energies are calculated from a renormalizable
part of a non-renormalizable theory, hence dimensional regularization
can be used to calculate the general vacuum polarization function
with two fermions of different masses circulating in the loop \cite{fcmew,silva}.

\section{Numerical constraints from precision tests}

There are 3 free parameter in the model to confront with experiment.
These can be chosen the three dimensionful four-fermion couplings
$\lambda_{1,\,2,\,3}$, or more practically the two physical neutral
masses $M_{1}$, $M_{2}$ and the mixing angle, $c^{2}=\cos^{2}\phi$.
For the cutoff $\Lambda\simeq3$ TeV there is a maximum value for
the masses, $M_{1}\leq240$ GeV. $c^{2}$ has an upper bound depending
on the mass $M_{1}$, see Figure 5. The mass of the charged fermion
is given by the solution of the gap equations, the value of $M_{+}$
is close to, but not equal to $c^{2}M_{1}+s^{2}M_{2}$.

If there is no real mixing $c^{2}=0$; or if $M_{1}=M_{2}=M_{+}$,
then there is one degenerate vector-like fermion doublet and a decoupled
singlet, and ${S}$ and ${T}$ vanish explicitely. In this case the
new sector does not violate $SU_{L}(2)$ and there is an exact custodial
symmetry. Increasing the mass difference in the remnants of the original
doublet by increasing the $\left|M_{1}-M_{2}\right|$ mass difference
and/or moving away from the non-mixing case $c^{2}=0,$ we get a higher
${S}$ and ${T}$. For small violation of the symmetries ${S}$ and
${T}$ are expected to be small. Numerical evaluation shows that for
the new masses in the range allowed by the LEP bound, gap equations
and unitarity the $U$ parameter is indeed an order of magnitude smaller
than the $T$ parameter and generally smaller than $S$. $U$ is always
in the experimental window. In case of relatively small masses the
oblique parameters are understood as rough estimates, but still in
agreement with experiment.

\begin{figure}[t]
\begin{centering}
\includegraphics[scale=0.5]{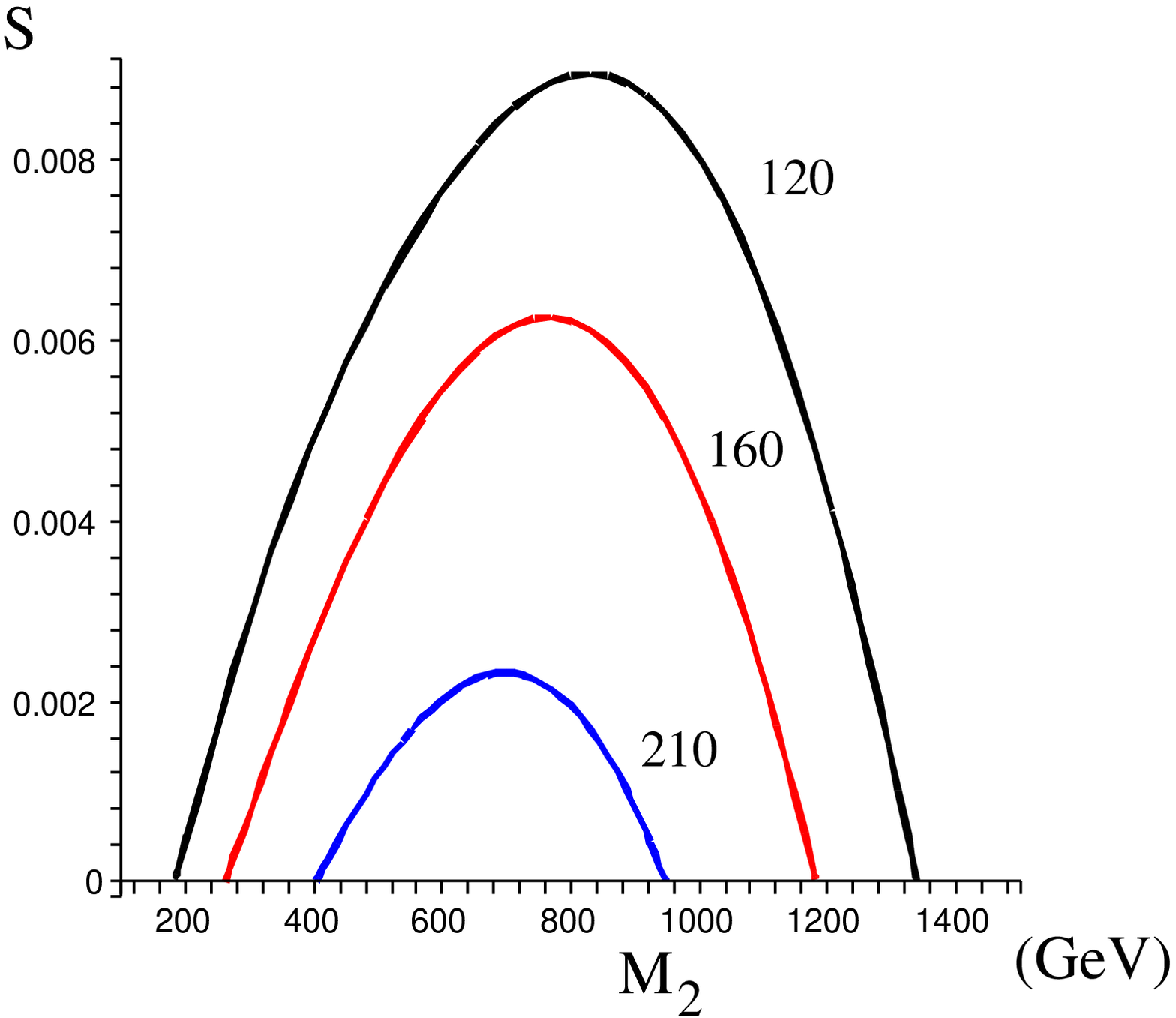} 
\par\end{centering}

\centering{}Figure 8. The maximum value of the ${S}$ parameter vs.
$M_{2}$ for $M_{1}=120,\,160,\,210$ GeV. The 95 \% C.L. bounds {[}-0.296,
0.096{]} are outside the figure.
\end{figure}

Generally the $S$ parameter depends only on the masses of the new
particles and the mixing angle. For the solutions of the gap equations
fulfilling perturbative unitarity the $S$ parameter is always positive
and far below the 95 \% C.L. For a given $M_{1},\, M_{2}$ $S$ increases
with increasing $c^{2}$ and maximal for the highest $c^{2}$. This
maximum value of the $S$ parameter is plotted against $M_{2}$ for
three given $M_{1}$ in Figure 8. The small value of $S$ does not
constrain the parameters of the model.

The value of the $T$ parameter is always positive. The ${T}$ parameter
(\ref{eq:tpar}) sensitive to the differences and ratios of the masses
$M_{1,\,2,\,+}$. $T$ still varies for a given $(M_{1},M_{2})$ pair
depending on $M_{+}$ or equally on $c^{2}$; $T$ is maximal for
largest mass difference, for the largest $c^{2}$ allowed by the gap
equations and perturbative unitarity. The $T$ parameter can always
be in agreement with experiment for any $(M_{1},M_{2})$ pair for
small mixing, for $c^{2}=0$ the $T$ parameter vanishes identically.
We plotted the worst case in the $(M_{1},M_{2})$ plane, the possible
maximum value of the $T$ parameter; it is given by the maximum $M_{2}-M_{+}$
mass difference or equally for maximal $c^{2}$.

If the Higgs is heavy, e.g. $M_{H}=300$ GeV (\ref{Sexp}, \ref{Texp})
the central value of ${S}$ decreases and ${T}$ increases compared
to the light Higgs case. The ${S}$ parameter still in agreement with
the predictions of the model. Incrasing the Higss mass the Standard
Model moves away in the (S,T) plane from the experimentally allowed
ellipse, see \cite{LEPEWWG}. The negative contribution ($-.09$)
of the heavy Higgs to the ${T}$ parameter can be compensated by the
positive ${T}$ contribution of the new fermions with considerable
mass difference. For example (160, 800) GeV and the largest mixing
$c^{2}\sim0.115$ allowed by the gap equations and unitarity gives
$\Delta T\simeq0.1$. Even heavier Higgs boson can be compensated
as can be read off from Figure 9. Non-degenerate vector-like fermions
with reasonable mixing allow a space for heavy Higgs in the precision
tests of the Standard Model.

\begin{center}
\begin{figure}
\begin{centering}
\includegraphics[scale=0.5]{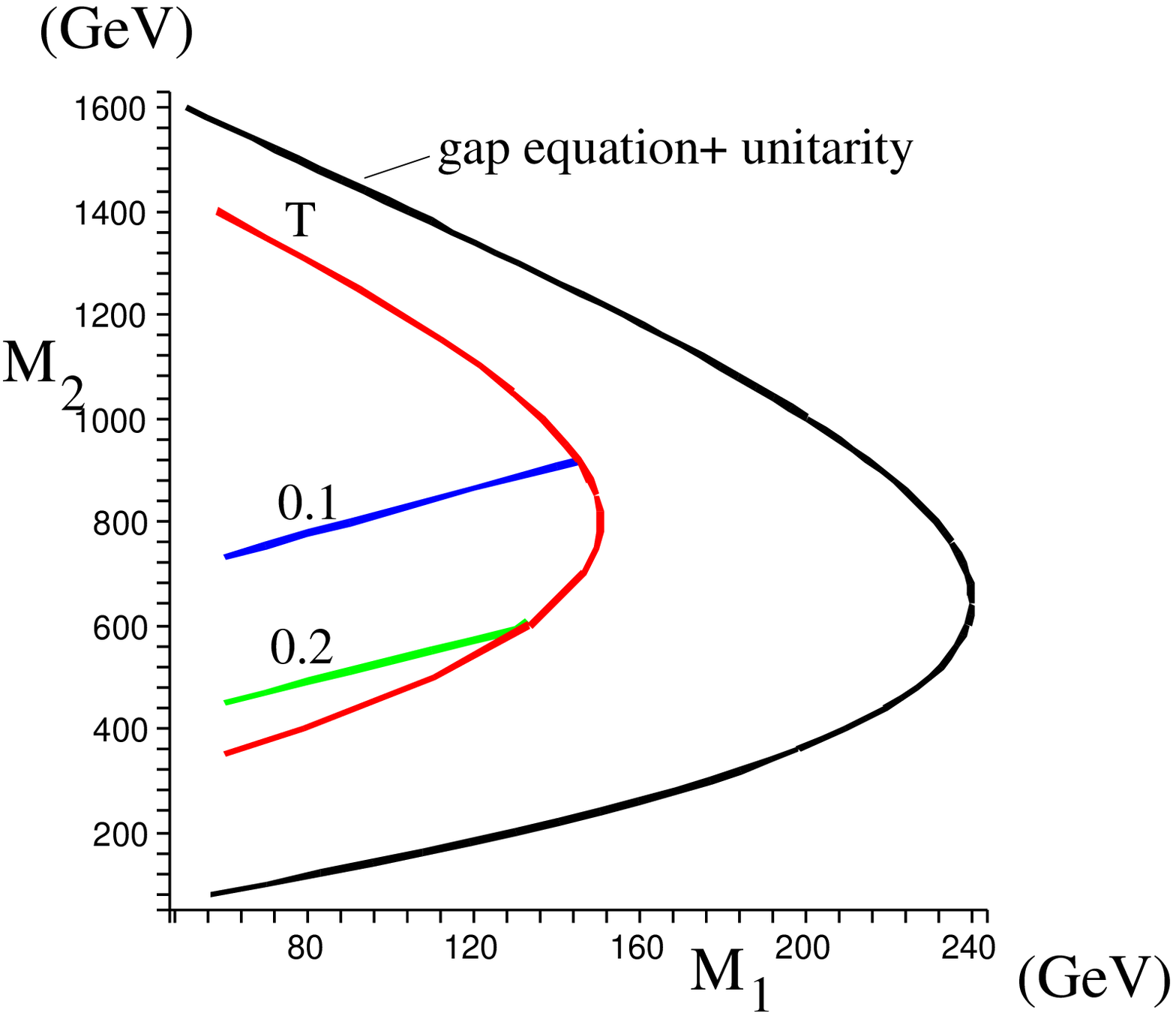} \includegraphics[scale=0.5]{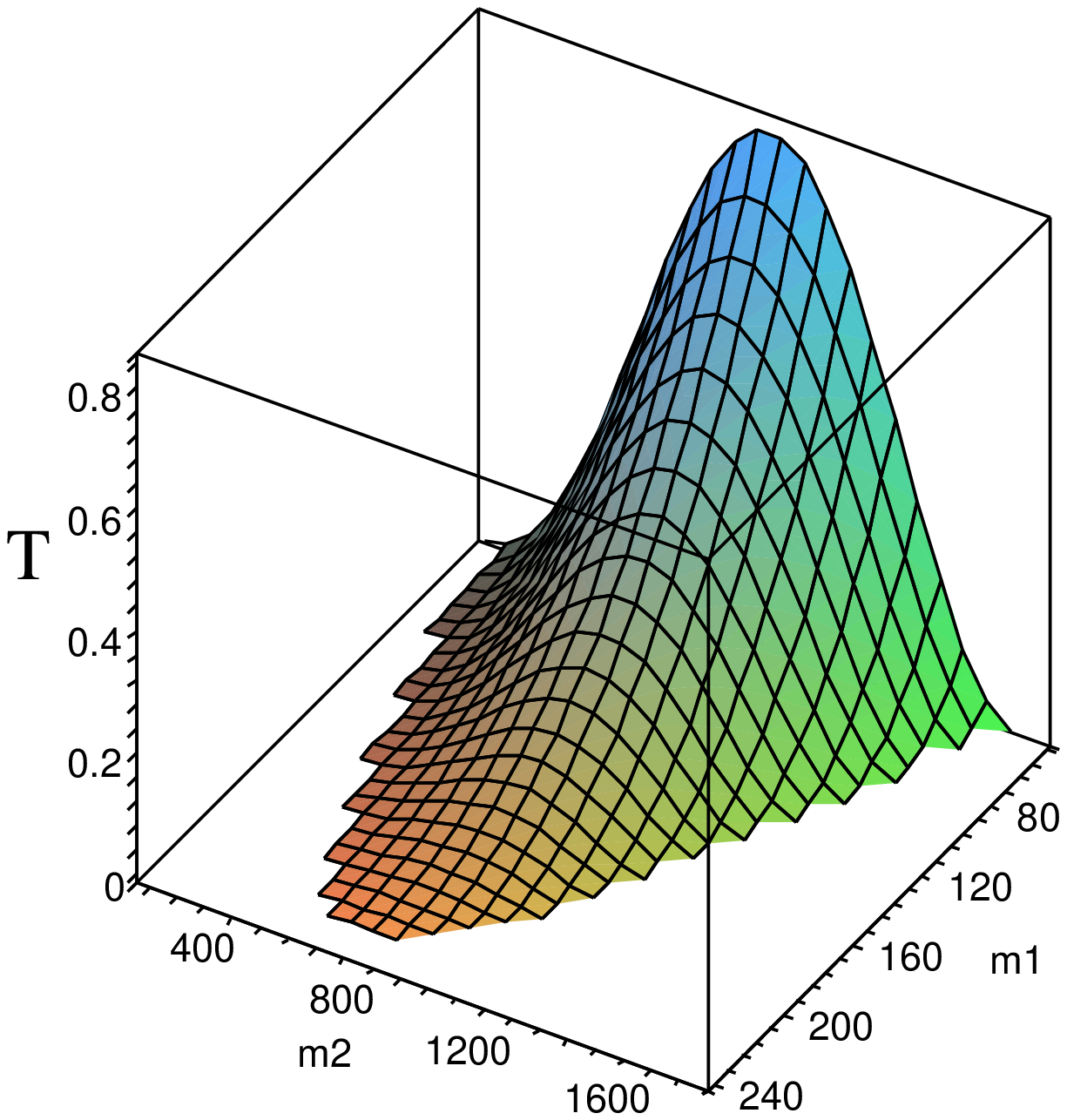} 
\par\end{centering}

Figure 9. Constraints on the $(M_{1},M_{2})$ plane. The solution
of the gap equations respecting perturbative unitarity are inside
the outer curve. The inner curve shows the region, where the ${T}$
parameter gives the maximum value of $c^{2}$ at 95 \% C.L.. Below
the 0.1 (blue) and 0.2 (green) line $c^{2}$ can exceed 0.1 and 0.2.
The right panel shows the maximum value of $T$ vs. $(M_{1},M_{2})$. 
\end{figure}

\par\end{center}

\section{Collider signatures}

In this section we study the production of the new fermions at LHC
and the planned linear collider. We focus on the preoduction of the
new charged femions with mass $M_{+}$, we denote it by $D^{+}$ and
its antiparticle by $D^{-}$.

Since the light standard fermions are coupled very weakly to the new
fermions producing pairs of new fermions is expected to be more considerable
from virtual $\gamma$ and Z exchanges, that is we consider the Drell-Yan
mechanism, $p^{(}\overline{p}^{)}\rightarrow D^{+}D^{-}+X$ via quark-antiquark
annihilation . The new fermion can only be produced in pairs because
of the $Z_{2}$ symmetry of the original Lagrangian \ref{eq:4fermion}.

The Drell-Yan cross section for the above hadronic collisions can
be written as \begin{eqnarray}
\sigma(p^{(}\overline{p}^{)}\rightarrow D^{+}D^{-}+X)=\int_{\tau_{0}}^{1}\, d\tau\int_{\tau}^{1}\,\frac{dx}{2x}\sum_{i}\sigma(q_{i}\overline{q}_{i}\rightarrow D^{+}D^{-})\cdot\nonumber \\
\left(f_{i}^{1}(x,\hat{s})f_{\bar{i}}^{2}(\tau/x,\hat{s})+f_{\bar{i}}^{1}(x,\hat{s})f_{i}^{2}(\tau/x,\hat{s})\right),\label{eq:sigmahadron}\end{eqnarray}
 where $x$ and $\tau/x$ are the parton momentum fractions, $\hat{s}=\tau s$
is the square of the centre of mass energy of $q_{i}\bar{q}_{i}$,
s is the same for the hadronic initial state, $f_{i}^{1}(x,\hat{s})$
means the number distribution of $i$ quarks in hadron 1 at the scale
$\hat{s}$ and the sum runs over the quark flavours u,d,s,c. In the
computation the MSTW parton distribution functions \cite{mstw} were
used. 

\begin{figure}
\begin{centering}
\includegraphics[scale=0.45]{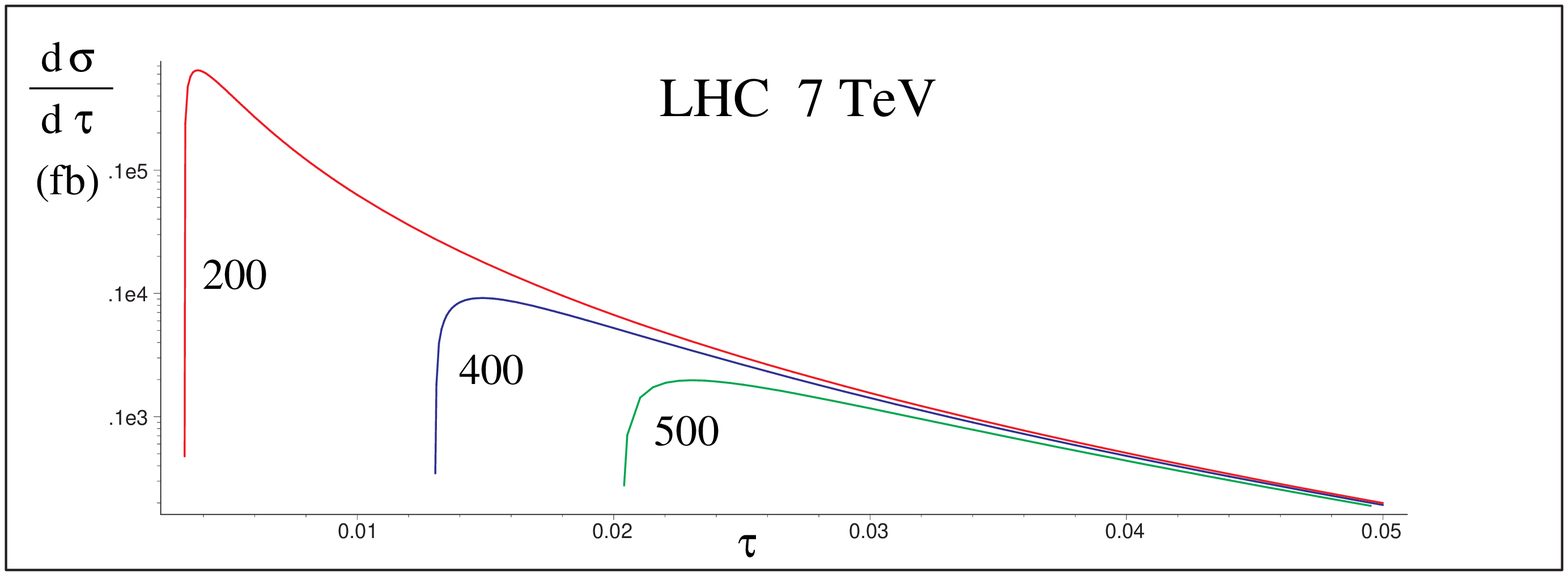}
\par\end{centering}

\centering{}Figure 10. The differential Drell-Yan production cross
section of $\Psi_{D}^{-}\Psi_{D}^{+}$ at the 7 TeV LHC.
\end{figure}

The angle integrated, colour averaged annihilation cross section $\sigma(q_{i}\overline{q}_{i}\rightarrow D^{+}D^{-})$
is calculated at the lowest order in the gauge couplings, and QCD
corrections are neglected. We hope this approximation shows the order
of magnitude of the cross section. We give the result of the charged
final state as there is no unknown mixing angle in the estimates.
The $D^{+}D^{-}$ pairs appear via $\gamma+Z$ exchange, the relevant
interactions are in (\ref{eq:4fermion}). The cross section at the
parton level is similar to the $\sigma(q_{i}\overline{q}_{i}\rightarrow\mu^{+}\mu^{-})$
cross section with increased masses (case of fourth family lepton),
see Figure 10.

The total cross sections for different masses are shown in Table 2,
and the expected number of events are very low at the delivered integrated
luminosity 35 pb$^{-1}$.

\begin{table}[h]
\begin{centering}
\begin{tabular}{c|c|c|c|}
$M_{+}$(GeV) & 200 & 400 & 500\tabularnewline
\hline
$\sigma$ (fb) & 215 & 9.3 & 2.6\tabularnewline
\end{tabular}
\par\end{centering}

\caption{Total production Drell-Yan cross section of $D^{+}D^{-}$ at the 7
TeV LHC.}

\end{table}

The new charged fermion $D^{+}$ may leave a charged track or a misplaced
vertex if it decays in a very short time to the lighter neutral new
fermion $\Psi_{1}$. Finally the lighter neutral fermion expected
to disappear leaving back missing energy and momentum, making it difficult
to select this model from other sources of dark matter candidates.
If new vector-like fermions can mix with the standard femions and
decay to standard particles one can search the new particles in jetmass
distributions \cite{jetmass} and can cope with the huge background.
We expect a higher yield at the 10-14 TeV LHC with the high design
luminosity.

A cleaner signal is expected at the next generation of linear collider.To
test the model at the forthcoming accelerators we consider the productions
of new fermion pairs in electron-positron annihilation. It is most
useful to investigate the case of a charged new fermion pair, we denote
this $D^{+}D^{-}$.

The contact graph from (\ref{eq:melectron}) yields the cross section 

\begin{equation}
\sigma\left(e^{+}e^{-}\rightarrow D^{+}D^{-}\right)=\frac{g_{e}^{2}}{16\pi}s\sqrt{1-4\frac{m_{+}^{2}}{s}}\left(1-\frac{5}{2}\frac{m_{+}^{2}}{s}\right),\label{eq:sigmaee}\end{equation}
where $s$ is the centre of mass energy squared. The cross section
is negligible at moderate $s$. For example at $h\sim\left(2TeV\right)^{-4}$,
$\sqrt{s}=1$TeV it is still at the order of $10^{-13}$ fb.

We expect a higher number of events from the photon and Z exchange
processes $e^{+}e^{-}\rightarrow\gamma,Z\;\rightarrow D^{+}D^{-}\!.$
The usual Standard Model coupling at the $e^{+}e^{-}Z$ vertex is\[
i\frac{g}{2\cos\theta_{W}}\gamma_{\mu}\left(g_{V}+\gamma_{5}g_{A}\right),\;\mathrm{where}\quad g_{V}=-\frac{1}{2}+2\sin^{2}\theta_{W},\; g_{A}=-\frac{1}{2}.\]
\begin{figure}
\begin{centering}
\includegraphics[scale=0.4]{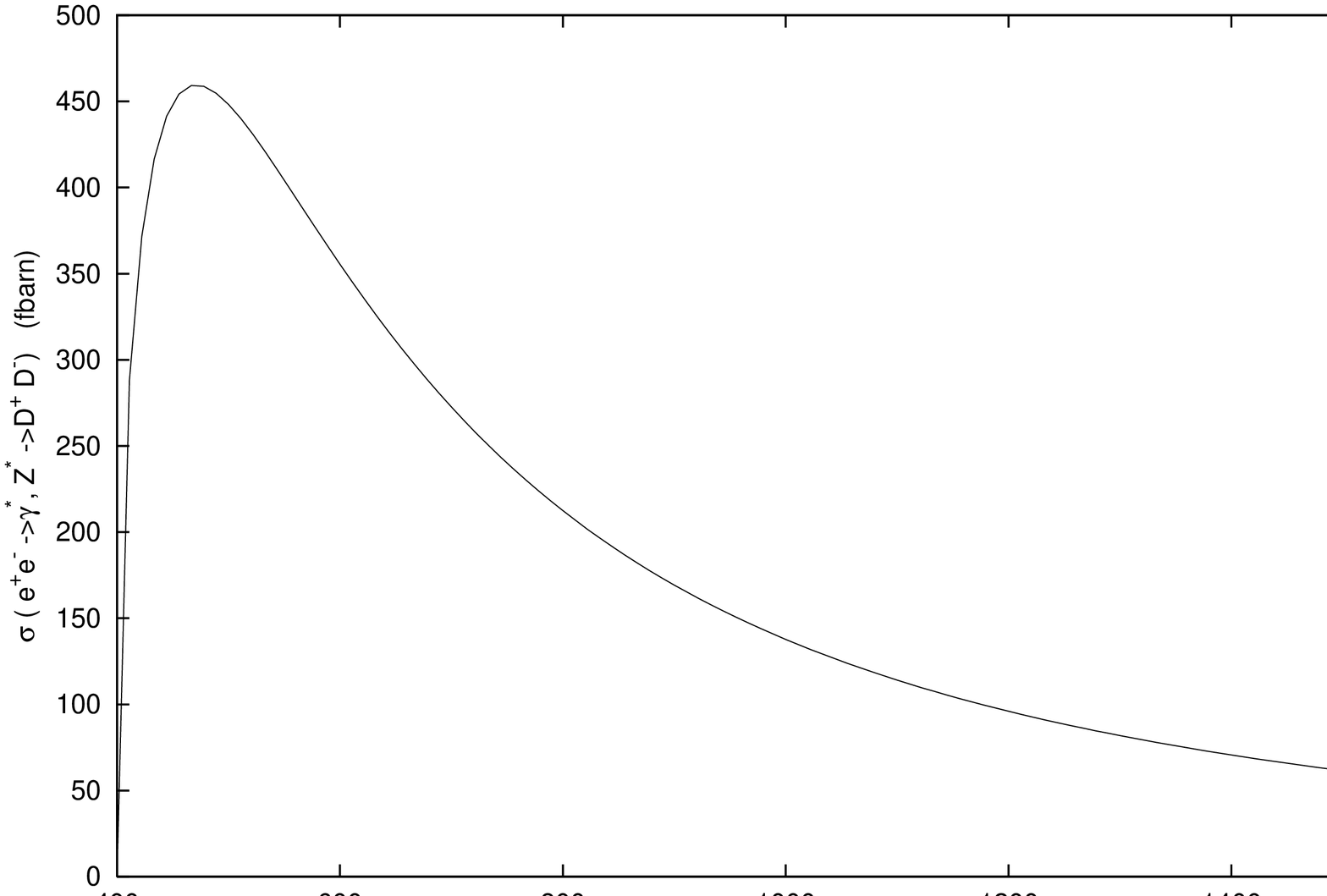}
\par\end{centering}

Figure 11. Cross section of $D^{+}D^{-}$ production at electron-positron
collider vs. $\sqrt{s}$ for $m_{+}=200$GeV
\end{figure}
\\
By making use of (\ref{eq:Llmix}) one obtains the cross section\begin{eqnarray*}
\sigma\left(e^{+}e^{-}\rightarrow D^{+}D^{-}\right)=\frac{1}{16\pi}\sqrt{1-4\frac{m_{+}^{2}}{s}}\frac{1}{s}\left|M\right|^{2} & ,\end{eqnarray*}
\begin{eqnarray}
\left|M\right|^{2} & = & \phantom{+}\frac{4}{3}e^{4}\frac{s+2m_{+}^{2}}{s}+\frac{2}{3}\frac{e^{4}}{\sin^{2}\theta_{W}\cos^{2}\theta_{W}}g_{V}\frac{s+2m_{+}^{2}}{s-m_{Z}^{2}}+\nonumber \\
 &  & +\frac{1}{12}\frac{e^{4}}{\sin^{4}\theta_{W}\cos^{4}\theta_{W}}\left(g_{V}^{2}+g_{A}^{2}\right)s\frac{s+2m_{+}^{2}}{\left(s-m_{Z}^{2}\right)^{2}},\label{eq: m2}\end{eqnarray}
where the three terms in $\left|M\right|^{2}$ are coming from photon
exchange , photon-Z interference and pure Z exchange. Similar cross
section belongs to the neutral pair productions, too. The cross section
rises fast after the threshold, at high energies it falls off as $1/s$
reflecting that all the interactions are renormalizable in the process.%
\begin{table}[h]
\begin{centering}
\begin{tabular}{|l|c|c|c|}
\hline 
$m_{+}$( GeV)  & 100 & 150 & 200\tabularnewline
\hline
$\sigma\left(e^{+}e^{-}\rightarrow D^{+}D^{-}\right)$ (fb) & 560 & 535 & 450\tabularnewline
\hline
\end{tabular}
\par\end{centering}

\caption{Cross section of $D^{+}D^{-}$ production at $\sqrt{s}=$500 GeV}

\end{table}
 The cross section is given in Table 3. for a few masses and plotted
versus $\sqrt{s}$ in Fig. 11. for $M_{+}=200$ GeV. At a linear collider
of $\sqrt{s}=500$GeV (TESLA) and integrated luminosity 50 $fb^{-1}$/year
a large number of events is expected. 

\begin{table}[h]
\begin{centering}
\begin{tabular}{|l|c|c|c|c|}
\hline 
$m_{+}$( GeV)  & 100 & 200 & 400 & 700\tabularnewline
\hline
$\sigma\left(e^{+}e^{-}\rightarrow D^{+}D^{-}\right)$ (fb) & 62 & 61 & 60 & 32\tabularnewline
\hline
\end{tabular}
\par\end{centering}

\caption{Cross section of $D^{+}D^{-}$ production at $\sqrt{s}=$1500 GeV}

\end{table}

The cross section at $\sqrt{s}=1500$GeV is an order of magnitude
smaller but with an integrated luminosity of 100 $fb^{-1}$ per annum
a large number of events appears and higher mass range can be searched
for.

\section{Conclusion}

In this chapter we have investigated a new dynamical symmetry breaking
model of the electroweak symmetry based on four-fermion interactions
of new hypothetical doublet and singlet vector-like fermions. Four-fermion
interactions are postulated involving the fermions and the standard
and new fermions. Gap equations were derived and we have found the
conditions for dynamical symmetry breaking, in the vacuum non-diagonal
condensates are formed. The lightest new particle is neutral and perturbative
unitarity sets an upper bound for its mass depending on the cutoff.
This particle is an ideal dark matter candidate. In the low energy
effective theory limit the Higgs is a composite particle. The $S$
and $T$ oblique parameters were calculated and presented. The solutions
of the gap equations provide masses that are always in the experimental
window of the $S$ parameter. The $T$ parameter measures the deviation
from custodial symmetry. The experimental data gives an upper bound
for the mixing angle, but there is always a room for this type of
new physics. This alternative of the Standard Model nicely accommodates
a composite heavy Higgs in the precision electroweak test of the Standard
Model. The vector-like quarks can easily compensate the negative contribution
of a heavy Higgs invalidating the light Higgs preference of the present
precision tests. We have presented the Drell-Yan cross section for
the production of the new charged fermion at the 7 TeV LHC, the expected
number of events is rather small with the 35 pb$^{-1}$ luminosity
delivered in 2010. The cross sections for linear electron-positron
colliders are higher and are more promising for a potential discovery.
Vector-like fermions appear in several researches beyond the Standard
Model physics and can elegantly accomodate a heavy Standard Model
like Higgs and provide a competitive dark matter candidate.

\subsubsection*{ACKNOWLEDGEMENT}

The authors dedicate this chapter to the late George Pócsik for collaboration
on the early phases of this work.

\section*{Appendix A. Regularization with momentum cutoff}

There are low energy theories, like the fermion condensate model,
which have an intrinsic cutoff, i.e. the upper bound of the model.
The naive calculation of divergent Feynman graphs with a momentum
cutoff is thought to break continuous symmetries of the model. In
this case the gauge invariance of the two point function with two
different fermion masses in the loops can be reconstructed by subtractions
leading to finite ambiguity. To avoid these problems we used dimensional
regularization in $d=4-2\epsilon$ and identified the poles at $d=2$
with quadratic divergencies while the poles at $d=4$ with logarithmic
divergencies \cite{hagiwara}. Carefully calculating the one and two
point Passarino-Veltman functions in the two schemes the divergencies
are the following in the momentum cutoff regularization \begin{eqnarray}
4\pi\mu^{2}\left(\frac{1}{\epsilon-1}+1\right) & = & \Lambda^{2},\\
\frac{1}{\epsilon}-\gamma_{E}+\ln\left(4\pi\mu^{2}\right)+1 & = & \ln\Lambda^{2},\end{eqnarray}
 where $\mu$ is the mass-scale of dimensional regularization. The
finite part of a divergent quantity is defined by \begin{equation}
f_{{\rm finite}}=\lim_{\epsilon\rightarrow0}\left[f(\epsilon)-R(1)\left(\frac{1}{\epsilon-1}+1\right)-R(0)\left(\frac{1}{\epsilon}-\gamma_{E}+\ln4\pi+1\right)\right],\end{equation}
 where $R(1)$,  are the residues of the poles at $\epsilon=1,\,0$
respectively.

We have found that contrary to the expectations the ambiguity of the
cutoff regularization scheme is coming from the replacement of \begin{equation}
l_{\mu}l_{\nu}\rightarrow g_{\mu\nu}l^{2}/4\label{eq:negyed}\end{equation}
 and not from shifting the loop-momentum $(l)$ .

In \cite{new} we have worked out a symmetry preserving regularization
in four dimensions. The key point is that tracing and divergent integration
are not commutative. Under divergent integrals regulated by momentum
cutoff in the new method the following identification will respect
gauge and Lorentz symmetry during the calculation

\begin{equation}
\int_{\Lambda\: reg}d^{4}l_{E}\frac{l_{E\mu}l_{E\nu}}{\left(l_{E}^{2}+m^{2}\right)^{n+1}}:=\frac{1}{2n}g_{\mu\nu}^{(E)}\int_{\Lambda\: reg}d^{4}l_{E}\frac{1}{\left(l_{E}^{2}+m^{2}\right)^{n}},\ \ \ \ \ n=1,2,...\label{eq:idn}\end{equation}
This identification is Lorentz invariant, in gauge theories \eqref{eq:idn}
guarantees the validity of the Slavnov-Taylor identities. It is shown
in \cite{anom} that the ABJ triangle anomaly can be correctly calculated
with this regularization.


\begin{thebibliography}{55}
\bibitem{pdg} K. Nakamura \emph{et al} (Particle Data Group) 2010
J. Phys. G.: Nucl. Part. Phys. \textbf{37} 0075021 .

\bibitem{tevatron10}CDF and D0 Collaboration,   
Phys. Rev. Lett. 104 (2010) 061802, updated in arXiv:1007.4587 [hep-ex].

\bibitem{gfitter} M.~Goebel,   
PoS {\bf ICHEP2010} (2010) 570,   arXiv:1012.1331 [hep-ph].

\bibitem{Weinberg}S.Weinberg, Phy. Rev. D \textbf{13}, 974 (1976).

\bibitem{Weinberg2}S.Weinberg, Phy. Rev. D \textbf{19}, 1277 (1979).

\bibitem{Susskind}L.Susskind, Phy. Rev. D 20, 2619 (1979).

\bibitem{Hillpr} C.~T.~Hill and E.~H.~Simmons,   
Phys.\ Rept.\  {\bf 381} (2003) 235,   [Erratum-ibid.\  {\bf 390} (2004) 553].

\bibitem{etc} E.~Eichten and K.~D.~Lane, 
  Phys.\ Lett.\  B {\bf 90} (1980) 125.

\bibitem{etc2} S.~Dimopoulos and L.~Susskind,   
 Nucl.\ Phys.\  B {\bf 155} (1979) 237.

\bibitem{holdom}B.~Holdom, 
Phys.\ Rev.\  D {\bf 24} (1981) 1441.

\bibitem{bando}K.~Yamawaki, M.~Bando and K.~Matumoto,   
Phys.\ Rev.\ Lett.\  {\bf 56} (1986) 1335.

\bibitem{latticeconf}T.~Appelquist, G.~T.~Fleming and E.~T.~Neil,  
Phys.\ Rev.\ Lett.\  {\bf 100} (2008) 171607.

\bibitem{sannino}  F.~Sannino and K.~Tuominen,   
Phys.\ Rev.\  D {\bf 71} (2005) 051901.

\bibitem{t1}W.~A.~Bardeen,C.T.~Hill and M.~Lindner, Phys.Rev. D \textbf{\textit{41}} 1647 (1990).

\bibitem{t2} C.T. Hill, Phys.Lett. \textbf{\textit{B266}}, 419 (1991).

\bibitem{t3} H.~C.~Cheng, B.~A.~Dobrescu and C.~T.~Hill,   
Nucl.\ Phys.\  B {\bf 589} (2000) 249.

\bibitem{t4} Y.~Bai, M.~Carena and E.~Ponton, 
Phys.\ Rev.\  D {\bf 81} (2010) 065004.   

\bibitem{lH} N. Arkani-Hamed, A.G. Cohen and H. Georgi, Phys.Lett.
\textbf{\textit{B513}}, 232 (2001).

\bibitem{kishier}L.~Giusti, A.~Romanino and A.~Strumia,
Nucl.\ Phys.\  B {\bf 550} (1999) 3.

\bibitem{lh1} N. Arkani-Hamed, A.G. Cohen, T. Gregoire and J.G. Wacker,
JHEP \textbf{\textit{0208}}, 020 (2002).

\bibitem{lh}N. Arkani-Hamed, A.G. Cohen, E. Katz, A.E. Nelson,T.
Gregoire, Jay G. Wacker, JHEP \textbf{0208}, 021 (2002).

\bibitem{pshiggs}H. Georgi and D. Pais, Phys. Rev. \textbf{D10} 539
(1974), ibid \textbf{D12} 508 (1975).

\bibitem{comp}D.~B.~Kaplan, H.~Georgi and S.~Dimopoulos,   
Phys.\ Lett.\  B {\bf 136}, 187 (1984).

\bibitem{compmin} K.~Agashe, R.~Contino and A.~Pomarol,   
Nucl.\ Phys.\ B {\bf 719} (2005) 165.

\bibitem{csaki} C.~Csaki, C.~Grojean, H.~Murayama, L.~Pilo and
J.~Terning, 
.\prd{46,}{1992}{381}.

\bibitem{fcm}G. Cynolter, E. Lendvai and G. Pócsik, Eur. Phys. J.
\textbf{C46}: 545 (2006).

\bibitem{UED}T.~Appelquist, H-C.~Cheng, B. A.~Dobrescu, 
\prd{64}{2001}{035002}.  

\bibitem{improved}Riccardo Barbieri, Lawrence J. Hall, Vyacheslav
S. Rychkov, Phys. Rev. \textbf{D74}: 015007 (2006).

\bibitem{darkmatter}R. Enberg, P.J. Fox, L.J. Hall, A.Y. Papaioannou,
M. Papucci, JHEP \textbf{0711}: 014 (2007); Rakhi Mahbubani, Leonardo
Senatore, Phys. Rev. \textbf{D73}: 043510 (2006).

\bibitem{dmhiggs}F. D'Eramo, Phys.Rev.D76:083522 (2007). 

\bibitem{top}W.A. Bardeen, C.T. Hill and M. Lindner, Phys.Rev. \textit{D}\textbf{\textit{41}}
1647 (1990); C.T. Hill, Phys.Lett. \textit{B}\textbf{\textit{266}},
419 (1991); M. Lindner and D. Ross, Nucl.Phys. \textit{B}\textbf{\textit{370}},
30 (1992); Bogdan A. Dobrescu and Christopher T. Hill, Phys. Rev.
Lett. \textbf{81}, 2634 (1998).

\bibitem{njl}Y. Nambu and G. Jona-Lasinio, Phys. Rev. \textbf{122},
345 (1961); Y. Nambu and G. Jona-Lasinio, Phys. Rev. \textbf{124},
246 (1961)\textbf{.}

\bibitem{vcm2}G. Cynolter, E. Lendvai and G. Pócsik, Eur. Phys. J.
C\textbf{38}, 247 (2004).

\bibitem{vcm3}G.~Cynolter, E.~Lendvai and G.~Pocsik,
arXiv:hep-ph/0412285;
C.~P.~Hays, L.~Bruchers, R.~Santos, A.~Gutierrez-Rodriguez, M.~S.~Berger, G.~Cynolter and H.~N.~Long, ``Search for the Higgs Boson,'' ISBN 1-59454-861-7, 2006.


\bibitem{stu}M. E. Peskin and T. Takeuchi, Phys. Rev. D \textbf{46,}
381 (1992).

\bibitem{maek}N.~Maekawa, 
Phys.\ Rev.\  D {\bf 52} (1995) 1684.

\bibitem{maek2}N.~Maekawa, 
Prog.\ Theor.\ Phys.\  {\bf 93} (1995) 919.

\bibitem{fcmgap}G. Cynolter and E. Lendvai, J. Phys G \textbf{34},
1711 (2007).

\bibitem{cust}P. Sikivie \textit{et al}., Nucl. Phys. B \textbf{173}
189, (1980).

\bibitem{klev}S. P. Klevansky, Rev. Mod. Phys. \textbf{64}, No. 3
(1992).

\bibitem{unit}B.Lee, C.Quigg and H.Thacker, Phys. Rev. D \textbf{16},
1519 (1977); D.Dicus and V.Mathur, Phys. Rev. D \textbf{7}, 3111 (1973).

\bibitem{unitvcm}G. Cynolter, A. Bodor and G. Pocsik, Heavy Ion Phys.
\textbf{7}, 245 (1998).

\bibitem{unitfermion}T. Appelquist, Michael S. Chanowitz, Phys. Rev.
Lett. \textbf{59}, 2405 (1987), Erratum-ibid. \textbf{60},1589 (1988).

\bibitem{Pocsik}G.~P\'ocsik, E.~Lendvai and G.~Cynolter,  
Acta Phys.\ Polon.\  B {\bf 24} (1993) 1495.

\bibitem{lep2}R. Barbieri, A. Pomarol, R. Rattazzi and A. Strumia,
Nucl. Phys. B\textbf{703}, 127 (2004).

\bibitem{LEPEWWG}LEP Electroweak Working Group homepage, http://lepewwg.web.cern.ch/LEPEWWG.

\bibitem{stu6}I. Maksymyk, C.P. Burgess and David London, Phys.Rev.
D\textbf{50}, 529 (1994); G. Altarelli, R. Barbieri and S. Jadach,
Nucl. Phys. B\textbf{369} 3 (1992).

\bibitem{fcmlambda} G.~Cynolter, E.~Lendvai and G.~P\'ocsik,
Mod.\ Phys.\ Lett.\  A {\bf 24} (2009) 2331.

\bibitem{fcmew} G.~Cynolter and E.~Lendvai, 
 Eur.\ Phys.\ J.\ C \textbf{58}, 463 (2008).

\bibitem{silva}L. Lavoura and J. P. Silva, Phys. Rev. D\textbf{47},
2046 (1993).

\bibitem{mstw}A.~D.~Martin, W.~J.~Stirling, R.~S.~Thorne and G.~Watt,
Eur.\ Phys.\ J.\  C {\bf 63} (2009) 189.

\bibitem{jetmass}Witold Skiba and David Tucker-Smith, Phys.Rev. D\textbf{75}:115010
(2007). 

\bibitem{hagiwara}K.~Hagiwara, S.~Ishihara, R.~Szalapski and D.~Zeppenfeld,   
Phys.\ Rev.\  D {\bf 48} (1993) 2182.

\bibitem{new} G.~Cynolter and E.~Lendvai,   
arXiv:1002.4490 [hep-ph].

\bibitem{anom} G.~Cynolter and E.~Lendvai,   
arXiv:1012.4648 [hep-ph].
\end{thebibliography}
\end{document}